\documentclass[aps,preprint,floatfix,nofootinbib,showpacs]{revtex4-1}
\pdfoutput=1
\usepackage{dcolumn}% Align table columns on decimal point
\usepackage{bm}% bold math
\usepackage{graphicx}% Include figure files
\usepackage{amssymb,amsmath}
\usepackage{multirow}
\usepackage{units,changes}
\usepackage{color,url}
\usepackage{tabu}
\usepackage{array}
\usepackage[colorlinks=true,urlcolor=blue,anchorcolor=blue
,citecolor=blue,filecolor=blue,linkcolor=blue,menucolor=blue
,linktocpage=true,pdfproducer=medialab,pdfa=true]{hyperref}
\usepackage{colordvi}
\usepackage{subcaption}
\def\beqa{\begin{eqnarray}}
\def\eeqa{\end{eqnarray}}

\usepackage{soul}
\usepackage{hhline} 

\begin{document}

\title{Current Status of Inert Higgs Dark Matter with Dark Fermions}

\def\slash#1{#1\!\!\!/}
\author{Yi-Zhong Fan$^{1,2}$, Yao-Yu Li$^{1,2}$, Chih-Ting Lu$^3$, Xiao-Yi Luo$^3$, Tian-Peng Tang$^1$, Van Que Tran$^{4,5}$, Yue-Lin Sming Tsai$^{1,2}$}
\affiliation{$^1$Key Laboratory of DM and Space Astronomy, Purple Mountain Observatory, Chinese Academy of Sciences, Nanjing 210033, China}
\affiliation{$^2$School of Astronomy and Space Science, University of Science and Technology of China, Hefei, Anhui 230026, China}
\affiliation{$^3$Department of Physics and Institute of Theoretical Physics, Nanjing Normal University, Nanjing, 210023, China} 
\affiliation{$^4$Tsung-Dao Lee Institute \& School of Physics and Astronomy, Shanghai Jiao Tong University, Shanghai 200240, China}
\affiliation{$^5$Phenikaa Institute for Advanced Study, Phenikaa University, Yen Nghia, Ha Dong, Hanoi 100000, Vietnam}

\begin{abstract} 

The precision measurements of the muon magnetic moment and the $W$ boson mass have sparked interest in the potential deviations from standard model (SM) predictions. 
While it may be premature to attribute any excesses in these precision measurements to new physics, they do offer a valuable indication of potential directions for physics beyond the SM. 
Additionally, the particle nature of dark matter (DM) remains a crucial enigma. 
Despite the absence of any definitive DM signal in direct detection and collider experiments, 
the Galactic Center GeV $\gamma$-ray excess and the AMS-02 antiproton ($\overline{p}$) excess could potentially offer hints related to the evidence of DM. 
Motivated by these observations, we propose a simple DM model that addresses all these issues.
This model extends the SM by incorporating singlet and doublet Dirac fermion fields, along with a doublet complex scalar field. 
For the viable parameter regions in this model, 
we find that future upgrades of the Large Hadron Collider and DM direct detection experiments can only partially probe them, while future high-energy muon colliders hold promise for exploring the unexplored parameter space.
\end{abstract}

\maketitle 

\section{Introduction} 

Although the particle standard model (SM) has achieved remarkable success with countless experimental confirmations, two clouds still loom over it due to recent precision measurements of the muon magnetic moment and the $W$ boson mass. The former measurement was updated from the Fermilab experiment~\cite{Muong-2:2023cdq}, and the new world average was reported as $a_{\mu}^{\text{Exp}} = 116592059(22)\times 10^{-11}$, which deviates from the SM prediction by $5.1\sigma$ ($1.8\sigma$) when the hadronic vacuum polarization (HVP) is calculated using the data-driven method~\cite{Aoyama:2020ynm} (lattice calculation method~\cite{Borsanyi:2020mff}). The root cause of this tension in the HVP contribution from these two methods remains unsolved, prompting the question of whether new physics effects are involved in this discrepancy. The latter measurement was reported by the CDF collaboration at Fermilab with $m_{W,\text{CDF II}} = 80.4335\pm 0.0094$ GeV~\cite{CDF:2022hxs}, 
revealing about a $7\sigma$ deviation from the SM prediction ($m_{W,\text{SM}} = 80.360\pm 0.006$ GeV). In contrast, the ATLAS collaboration reported their $W$ boson mass measurement $m_{W,\text{ATLAS}} = 80.360\pm 0.016$ GeV~\cite{ATLAS:2023fsi}, and very recently the CMS collaboration reported $m_{W,\text{CMS}} = 80.3602 \pm 0.0099$ GeV~\cite{CMS:2024}, which is consistent with the SM prediction but as precise as the CDF II measurement. 
This discrepancy raises another puzzle, and further data are needed to confirm whether new physics effects are at play. Therefore, these excesses from the two precision measurements may unveil new physics beyond the SM, even without finding any evidence of new particles.

On the other hand, the dark matter (DM) problem remains unsolved. Strong astrophysical and cosmological evidence supports the existence of DM~\cite{Roos:2010wb}. Yet, apart from gravitational interaction, the particle nature of DM remains unknown. Among various DM candidates, Weakly Interacting Massive Particles (WIMPs)~\cite{Arcadi:2017kky,Bauer:2017qwy} are well-motivated and popular. WIMPs have masses around the electroweak scale, and their interaction strength with SM particles is expected to be close to the weak interaction. Due to their predictability, there are ongoing DM direct detection~\cite{Schumann:2019eaa}, DM indirect detection~\cite{Slatyer:2021qgc}, and collider experiments~\cite{Boveia:2018yeb} dedicated to searching for WIMPs. Although DM direct detection~\cite{LUX-ZEPLIN:2022xrq, XENONCollaboration:2023orw} and collider experiments~\cite{PerezAdan:2023rsl} have placed strong constraints on WIMPs, there are still some potential pieces of evidence from DM indirect detection. For example, the galactic center GeV excess (GCE)~\cite{Hooper:2010mq, Zhou:2014lva, Calore:2014xka, Daylan:2014rsa} and the AMS-02 antiproton ($\overline{p}$) excess~\cite{Cui:2016ppb, Cuoco:2016eej, Cui:2018klo, Cholis:2019ejx} provide indications that warrant further investigation.

Given the motivations mentioned above, there is an active demand for a model that can address these issues. We propose a simple DM model that extends the SM by introducing singlet and doublet Dirac fermion fields, along with a doublet complex scalar field. This model is designed to explain the $(g-2)_{\mu}$, the CDF $M_W$, as well as the GCE and AMS-02 $\overline{p}$ excesses.
We find the feasible parameter regions have the following intriguing features: (i) The mass of the scalar DM candidate is favored to lie within the range of $53$ to $74$ GeV. If the $W$ boson mass is consistent with the electroweak fit results, a DM mass heavier than $500$ GeV is still allowed. (ii) The compressed mass spectrum in the inert scalar sector is strongly constrained by current and upcoming LHC searches. (iii) The current constraints from the DM direct detection at LZ experiment already excluded a significant portion of the parameter space for the Higgs resonance. (iv) The neutral particles in  the leptophilic fermion sector are difficult to explore at the LHC; however, a high-energy muon collider could be an ideal machine for detecting them and testing the predictions of this model.

The structure of this paper is organized as follows. We first introduce our simple DM model in Sec.~\ref{sec:model}. We then discuss the extra $(g-2)_{\mu}$ and $M_W$ contributions, as well as the constraints in this model in Sec.~\ref{sec:th_exp}. Our main results are presented in Sec.~\ref{sec:result}. Finally, we conclude our findings in Sec.\ref{sec:conclude}.

\section{The model} 
\label{sec:model}

\begin{table}[t] 
\begin{center}
\begin{tabular}{c||c|c|c|c||c|c}\hline\hline  
&\multicolumn{4}{c||}{Fermion Fields} & \multicolumn{2}{c}{Scalar Fields} \\\hline
& ~$M_L$~ & ~$\mu_R$~ & ~$\Psi$~ & ~$\chi$~ & ~$H_1$~ &
 ~$H_2$~ \\\hline 
$SU(2)_L$ & $\textbf{2}$ & $\textbf{1}$ & $\textbf{2}$ & $\textbf{1}$ &
 $\textbf{2}$ & $\textbf{2}$ \\\hline 
$U(1)_Y$ & $-1/2$ & $-1$ & $-1/2$ & $0$ & $+1/2$& $+1/2$ \\\hline
$Z_2$ & $+$ & $+$ & $-$ & $-$ & $+$& $-$ \\\hline\hline
\end{tabular}
\caption{Contents of fermion and scalar fields
and their charge assignment under the $SU(2)_L\times U(1)_Y\times Z_2$ symmetry in this model. Here $M_L = \left(\nu_L, \mu^{-}_L \right)$ is the second generation of the lepton doublet and $\Psi = \left( \psi^0, \psi^- \right)$. 
}
\label{tab:fields} 
\end{center}
\end{table}
 
The SM is extended by introducing singlet and doublet Dirac fermion fields ($\chi$ and $\Psi$) as well as a doublet complex scalar field ($H_2$). To ensure the stability of the DM candidate, we impose a $Z_2$ symmetry on the model and assume that all new fields (SM fields) are $Z_2$ odd (even). The relevant fields under the $SU(2)_L \times U(1)_Y \times Z_2$ symmetry in this model are shown in Table~\ref{tab:fields}, where $H_1$ represents the SM scalar doublet, $M_L = \left(\nu_L, \mu^{-}_L \right)$ is the second generation of the lepton doublet, and $\Psi = \left( \psi^0, \psi^- \right)$. The renormalizable Lagrangian involving these new fields can be written as 
\begin{equation}
    {\cal L} = {\cal L}_F +{\cal L}_S +{\cal L}_{\text{mix}}, 
\end{equation}
where 
\begin{align} 
{\cal L}_F = & \overline{\Psi}\left({\:/\!\!\!\! D} - M_{\Psi}\right)\Psi + \overline{\chi}\left({\:/\!\!\!\! \partial} -M_{\chi}\right)\chi +\left( y_{H_1}\overline{\chi}\Psi\cdot H_1 + h.c.\right) \,, 
\label{eq:LF}
\end{align}  

\begin{align} 
{\cal L}_S = & \mid D_{\mu} H_1\mid^2 + \mid D_{\mu} H_2\mid^2 -\mu^2_1\mid H_1\mid^2 -\mu^2_2\mid H_2\mid^2 -\lambda_1 \mid H_1\mid^4 -\lambda_2 \mid H_2\mid^4 \notag \\ 
& -\lambda_3 \mid H_1\mid^2 \mid H_2\mid^2 -\lambda_4\left( H^{\dagger}_1 H_2\right)\left( H^{\dagger}_2 H_1\right) 
-\frac{\lambda_5}{2}\left[\left( H^{\dagger}_1 H_2\right)^2 + h.c.\right] \,, 
\label{eq:LS}
\end{align}  

\begin{align} 
\label{eq:Lmix}
{\cal L}_{\text{mix}} = & \kappa_M \overline{\chi}M_L\cdot H_2 +\kappa_{\mu}\overline{\Psi}\mu_R H_2 + h.c.\,. 
\end{align}  
Here $D_{\mu}$ is the covariant derivative in the SM, and the $SU(2)_L$ contraction $\Psi\cdot H_{\alpha}=\epsilon_{ab}\Psi_a H_{\alpha b}$ with $\epsilon_{ab} = (i\sigma_2)_{ab}$ and $\alpha = 1,2$. 
Note that $M_{\Psi}$, $M_{\chi}$, $\mu_i$ ($i = 1,2$) are parameters with the same dimension as mass and $y_{H_1}$, $\lambda_i$ ($i=1-5$), $\kappa_M$, $\kappa_{\mu}$ are dimensionless parameters. Furthermore, we assume all dimensionless parameters are real in this work\footnote{Note that the Yukawa-type couplings $y_{H_1}$, $\kappa_M$, and $\kappa_{\mu}$ in Eqs.~\ref{eq:LF} and~\ref{eq:Lmix} are generally complex.}.

Some comments on this model are as follows: 
\begin{itemize}
\item The inclusion of the third term in Eq.~(\ref{eq:LF}) aims to generate the chiral enhancement effect in the one-loop process to $\Delta a_{\mu}$. Without the presence of this chiral enhancement effect, it becomes challenging for the vector-like leptons to provide a simultaneous explanation for $\Delta a_{\mu}$ and to satisfy the collider constraints~\cite{Okada:2014qsa}. 
\item Both the doublets $H_2$ and $\Psi$ can contribute to the oblique parameters, $S$, $T$, and $U$~\cite{Peskin:1991sw}, thereby alleviating the strong restrictions on the parameter space of the inert 2HDM~\cite{Fan:2022dck}.
\item {The singlet fermion field $\chi$ in this model can be either Dirac-type or Majorana-type. However, if $\chi$ is Majorana-type, as discussed in Ref.~\cite{Ma:2006km,Cai:2017jrq}, the parameter space under consideration would result in excessively large neutrino masses via the one-loop process. Thus, from a phenomenological perspective, the $(g-2)_{\mu}$ excess provides a strong indication for treating the $\chi$ field as Dirac-type rather than Majorana-type in this study.}
\item ${\cal L}_F$ or ${\cal L}_S$ alone can represent a Lagrangian for a pure fermion or scalar DM model. However, neither by itself can provide a significant contribution to $\Delta a_{\mu}$. Therefore, we combine both of them with ${\cal L}_{\text{mix}}$ in this work. 
\item 
In the minimal model, ${\cal L}_{\text{mix}}$, we consider only the second generation of the lepton sector. However, it can be straightforwardly extended to include all three generations. The couplings of the dark fermion to $e$, $\mu$, and $\tau$ can be treated as free and independent parameters. This flexibility allows us to set the couplings to $e$ and $\tau$ small enough that they do not affect the results of this analysis.

\end{itemize}

After electroweak symmetry breaking (EWSB), the masses of scalar fields can be represented as 
\begin{align} 
& m^2_h = -2\mu^2_1 = 2\lambda_1 v^2, \notag \\ 
& m^2_S = \mu^2_2 +\frac{1}{2}\left( \lambda_3 +\lambda_4 +\lambda_5\right) v^2, \notag \\ 
& m^2_A = \mu^2_2 +\frac{1}{2}\left( \lambda_3 +\lambda_4 -\lambda_5\right) v^2, \notag \\ 
& m^2_{H^{\pm}} = \mu^2_2 +\frac{1}{2}\lambda_3 v^2 \notag \,, 
\end{align} 
and the mass matrix of singlet-doublet fermion fields is 
\begin{equation}
 \left( \begin{array}{cc} 
            M_{\chi} & y_{H_1} v/\sqrt{2} \\
            y_{H_1} v/\sqrt{2} & M_{\Psi} \end{array} \right). \, 
\end{equation} 
Similarly, after EWSB, the neutral components $\chi$ and $\psi^0$ mix giving the mass eigenstates $n_1$, $n_2$, 
\begin{equation}
 \left( \begin{array}{c}
                \chi \\ 
                \psi^0 \end{array} \right) 
= 
 \left( \begin{array}{cc} 
            \cos\alpha & -\sin\alpha \\
            \sin\alpha & \cos\alpha \end{array} \right )\,
 \left( \begin{array}{c}
                n_1 \\ 
                n_2 \end{array} \right), 
\end{equation}
with mass eigenvalues 
\begin{equation}
m_{n_1,n_2} = \frac{1}{2}\left( M_{\Psi}+M_{\chi}\mp\sqrt{(M_{\Psi}-M_{\chi})^2 +2 y^2_{H_1} v^2}\right), 
\end{equation}
and the rotation angle $\alpha$ is
\begin{equation}
\sin 2\alpha =\frac{\sqrt{2}y_{H_1}v}{m_{n_2}-m_{n_1}}.
\end{equation}
We assume that $n_2$ is heavier than $n_1$. Finally, the mass of charged fermion $\psi^{\pm}$ in the doublet field can be given as $m_{\psi^{\pm}} = M_{\Psi}$. We take $m_h = 125$ GeV and $v = 246$ GeV to eliminate $\mu_1$ and $\lambda_1$, so there are ten free parameters in this model left. Besides, as $\lambda_2$ primarily influences the self-interactions of inert scalars and loop corrections~\cite{Sokolowska:2011sb,Abe:2015rja,Banerjee:2019luv,Tsai:2019eqi}, neither of which significantly impacts the main findings of this study, we can conveniently assign $\lambda_2 = 10^{-2}$.

Subsequently, we introduce several key parameters in our analysis: the neutral mass splitting, $\Delta^0$, the charged mass splitting, $\Delta^{\pm}$, the coupling $\lambda_S$ between $h$ and a pair of $S$, and the disparity between two fermion mass parameters, $\Delta_{\Psi\chi}$ in the following:    
\begin{align} 
& \Delta^0 = m_A - m_S, \notag \\ 
& \Delta^{\pm} = m_{H^{\pm}} - m_S, \notag \\ 
& \lambda_S = \left( \lambda_3 +\lambda_4 +\lambda_5\right) /2, \notag \\ 
& \Delta_{\Psi\chi} = M_{\Psi} - M_{\chi} \notag \,. 
\end{align} 
Utilizing the definitions provided above, we establish the scanning intervals for the nine free model parameters as follows:
\begin{align} 
& 30\leq m_S/\text{GeV}\leq 100, \notag \\
& 10^{-3}\leq\Delta^0/\text{GeV}\leq 500, \notag \\
& 1\leq\Delta^{\pm}/\text{GeV}\leq 500, \notag \\
& -1.0\leq\lambda_S\leq 1.0, \notag \\
& -\sqrt{4\pi} < y_{H_1} < \sqrt{4\pi}, \notag \\ 
& -\sqrt{4\pi} < \kappa_M < \sqrt{4\pi}, \notag \\ 
& -\sqrt{4\pi} < \kappa_{\mu} < \sqrt{4\pi}, \notag \\ 
& 600\leq M_{\Psi}/\text{GeV}\leq 1500, \notag \\ 
& 1\leq\Delta_{\Psi\chi}/\text{GeV}\leq 500 \notag \,. 
\end{align} 
Here we impose the stipulation that $S$ as the DM candidate, ensuring that $m_S < m_{n_1, n_2}$. 
In some specific parameter space, the neutral particle $n_1$ can be regarded as the next-to-lightest dark particle, and its possible mass range is discussed in Sec.~\ref{sec:results}. 
The selection of the scanning ranges for $m_S$, $\Delta^0$, $\Delta^{\pm}$, and $\lambda_S$ adheres to the guidelines outlined in our previous investigations~\cite{Fan:2022dck,Tsai:2019eqi}. Additionally, for perturbative consistency, we enforce the bounds $\lvert y_{H_1}\rvert$, $\lvert\kappa_M \rvert$, and $\lvert\kappa_{\mu}\rvert$ to be less than $\sqrt{4\pi}$. Furthermore, to evade the constraints imposed by ATLAS and CMS slepton searches~\cite{ATLAS:2019lff,CMS:2020bfa}, we conservatively require the condition $M_{\Psi} > 600$ GeV.

\section{ Other constraints and extra $(g-2)_{\mu}$, $M_W$ contributions }
\label{sec:th_exp}

In this section, we will provide an overview of the relevant theoretical and experimental constraints in Sec.~\ref{sec:constraint}. Subsequently,  we will initially delve into the manner in which our model contributes to $\Delta a_{\mu}$ and $M_W$, as outlined in Sec.~\ref{sec:g2_mw}.

\subsection{Theoretical and experimental constraints}
\label{sec:constraint}

The relevant constraints for this model can be categorized into theoretical and experimental aspects. Regarding the theoretical constraints, we incorporate considerations of perturbativity, stability, and tree-level unitarity in the scalar sector, all of which are implemented using the \texttt{2HDMC}. 
As for the experimental constraints, we outline them as follows:  
\begin{itemize}
\item Production of inert scalars at the LEP \\ 
Firstly, it's important to note that no signals have been reported by LEP for processes like $W^{\pm}\rightarrow (SH^{\pm}, AH^{\pm}$), and $Z\rightarrow (SA, H^+H^-)$~\cite{ParticleDataGroup:2014cgo}. This leads to conditions such as $m_{S,A} +m_{H^{\pm}} > m_W$, $m_A +m_S > m_Z$, and $2m_{H^{\pm}} > m_Z$. For the heavier inert scalars, they can also be directly produced from processes like $e^+ e^-\rightarrow SA$ and $e^+ e^-\rightarrow H^+ H^-$. We closely follow the strategies outlined in Refs.~\cite{Lundstrom:2008ai,Blinov:2015qva,Tsai:2019eqi} to recast the constraints based on searches conducted by the OPAL collaboration~\cite{OPAL:2003nhx}.  
\item Higgs boson invisible and exotic decays \\ 
When the inert scalars have masses lower than half of $m_h$, certain decay modes become kinematically possible. These include the Higgs invisible decay, $h\rightarrow SS$, and the Higgs exotic decays, $h\rightarrow AA, H^+ H^-$\footnote{Given that the soft visible objects cannot be identified at the LHC, we classify $h\rightarrow AA, H^+H^-$ as Higgs invisible decays when $\Delta^{0,\pm}$ is less than about $2$ GeV~\cite{ATLAS:2019lng}.}. In our analysis, we consider BR$(h\rightarrow\text{invisible}) < 10.7\%$~\cite{ATLAS:2023tkt} and BR$(h\rightarrow\text{undetected}) < 19\%$~\cite{ATLAS:2020qdt} at $95\%$ C.L.
\item The $h\rightarrow\gamma\gamma$ signal strength \\ 
If the newly introduced charged particles can interact with the Higgs boson, the $h\rightarrow\gamma\gamma$ signal strength measurements can indirectly impose constraints on them. The $h\rightarrow\gamma\gamma$ signal strength measurements from the ATLAS (CMS) collaborations are reported as follows~\cite{ATLAS:2022tnm,CMS:2021kom}:
\begin{equation}
\mu_{\gamma\gamma} = 1.04^{+0.10}_{-0.09} (1.12\pm 0.09).
\end{equation}
In this model, only $H^{\pm}$ can contribute to the $h\rightarrow\gamma\gamma$ process~\cite{Swiezewska:2012eh}. 
\item Mono-jet and compressed mass spectra searches at the LHC \\ 
A common approach for searching for DM at the LHC involves investigating final states characterized by significant missing transverse energy associated with visible particles. In this model, the mono-jet signal comprises a pair of $S$ particles originating from the Higgs boson, a $n_1\overline{n_1}$ pair originating from the $Z$ and Higgs bosons, and then $n_1\rightarrow\nu_{\mu}S$ and $\overline{n_1}\rightarrow\overline{\nu_{\mu}}S$, accompanied by at least one high-energy jet. Similarly, if the mass difference between $A$ and $S$ is sufficiently small, a pair of $A$ particles produced through Higgs boson or $SA$ production via the $Z$ boson can contribute to the mono-jet signature. 
On the other hand, when the mass splittings ($\Delta^0$ and/or $\Delta^{\pm}$) are not large enough, searches focusing on compressed mass spectra become pivotal. In this model, compressed mass spectra searches can be performed for pairs of $SA$, $AH^{\pm}$, and $H^+H^-$, generated through $q\overline{q}$ fusion and VBF processes as well as $pp\rightarrow n_1\overline{n_1}$ with $n_1\rightarrow\nu_{\mu}S, \nu_{\mu}A, \mu^-H^+$ and $\overline{n_1}\rightarrow\overline{\nu_{\mu}}S, \overline{\nu_{\mu}}A, \mu^+H^-$ at the LHC. We reanalyze the current ATLAS mono-jet search~\cite{ATLAS:2021kxv} and SUSY compressed mass spectra search~\cite{CMS:2021edw} using \texttt{Madgraph5}~\cite{Alwall:2014hca}, \texttt{Pythia8}~\cite{Sjostrand:2014zea}, \texttt{Delphes3}~\cite{deFavereau:2013fsa} and \texttt{Madanalysis 5}~\cite{Dumont:2014tja}, as described in more detail in Ref.~\cite{Tsai:2019eqi}.  Please note that in the upcoming section, we will demonstrate that $n_2$ and $\psi^{\pm}$ are sufficiently heavy to evade constraints from the current LHC data.  
\item DM relic density \\ 
The dominant channels for DM annihilation in this model depend on the specific mass spectrum. In the Higgs resonance regions where $m_S\simeq m_h/2$, the annihilation process $SS\rightarrow b\overline{b}$ mediated by the Higgs boson exchange prevails. In the region where $m_S > m_h/2$, the dominant annihilation channels are $SS\rightarrow W^+W^{-(\ast)}$, involving interactions through a four-vertex interaction, $s$-channel Higgs boson exchange, and $t$-channel $H^{\pm}$ exchange. Additionally, co-annihilation processes play a role, such as $SA\rightarrow f\overline{f}$ for small $\Delta^0$ and $SH^{\pm}\rightarrow\gamma W^{\pm}$ for small $\Delta^{\pm}$. Beyond these DM annihilation or co-annihilation channels found in the inert 2HDM, there are also new annihilation channels of $SS\rightarrow\mu^+\mu^-$ via the $t$-channel $\psi^{\pm}$ exchange and $SS\rightarrow\nu\overline{\nu}$ via the $t$-channel $n_1$, $n_2$ exchange. To quantify these processes, we numerically solve the Boltzmann equation using \texttt{micrOMEGAs}~\cite{Belanger:2018ccd}, accounting for all possible annihilation and co-annihilation contributions. The relic density of DM is required to align with the PLANCK measurement~\cite{Planck:2015ica}: 
\begin{equation}
    \Omega_{\text{DM}}
h^2 = 0.1198\pm 0.0015.
\end{equation} 
\item DM direct detection \\ 
The null signals from DM direct detection experiments have placed stringent constraints on the scattering cross section between DM and nuclei. For instance, the spin-independent DM-nuclear scattering cross section has been excluded above $9.2\times 10^{-48}$ cm$^2$ for a DM mass of $36$ GeV at the $90\%$ confidence level from the current LZ experiment~\cite{LUX-ZEPLIN:2022xrq}. 
In this model, the leading-order contributions to DM-nuclear scattering originate from DM-quark/gluon elastic scattering mediated by the exchange of a Higgs boson in the $t$-channel. We utilize \texttt{micrOMEGAs} to compute the spin-independent DM-nucleon scattering cross section. 
\end{itemize}

\subsection{Contributions to $\Delta a_{\mu}$ and $M_W$} 
\label{sec:g2_mw}

\begin{figure}[h]
\centering 
\includegraphics[width=0.4\textwidth]{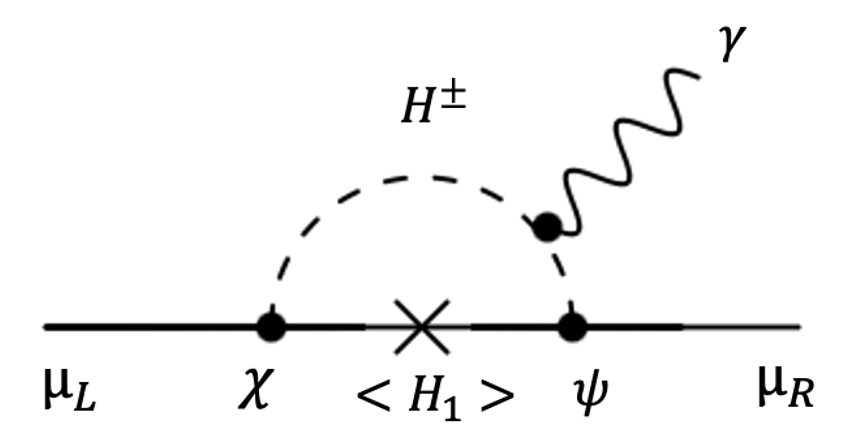}
\caption{The one-loop chirally enhanced Feynman diagram to the $\Delta a_{\mu}$ in this model.}
\label{fig:mu_g2}
\end{figure}

\begin{figure}[tb]
    % \centering
	\includegraphics[width=0.45\textwidth]{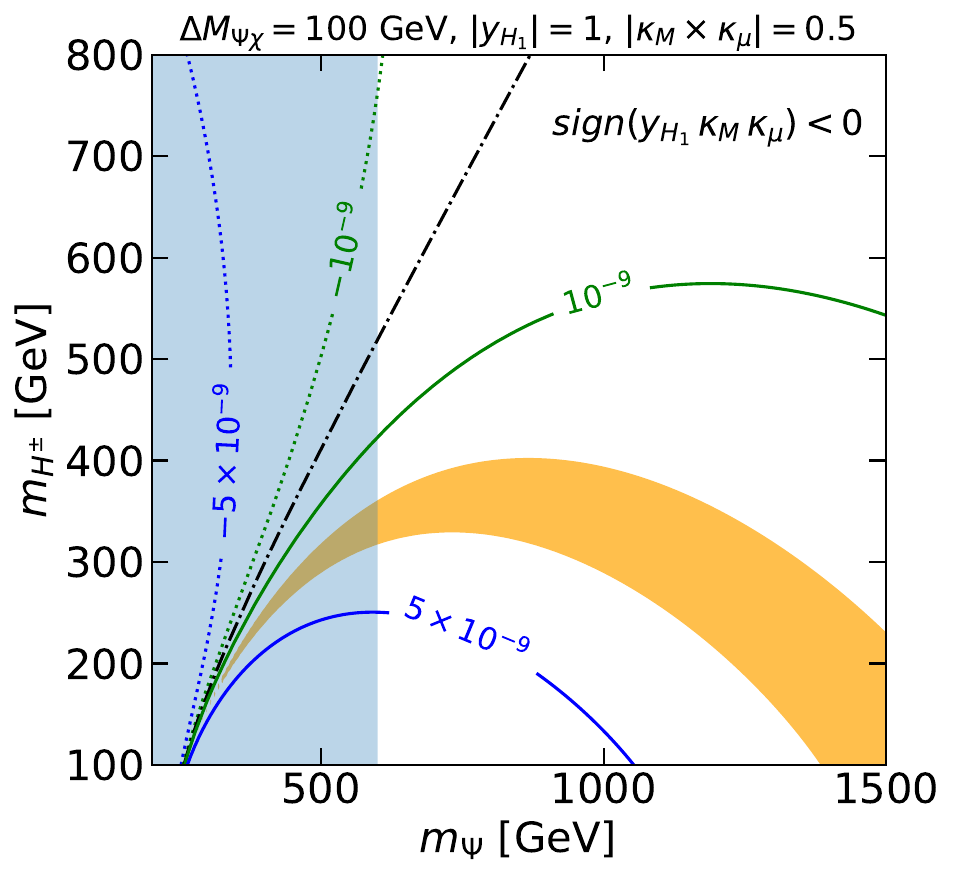}
    \includegraphics[width=0.45\textwidth]{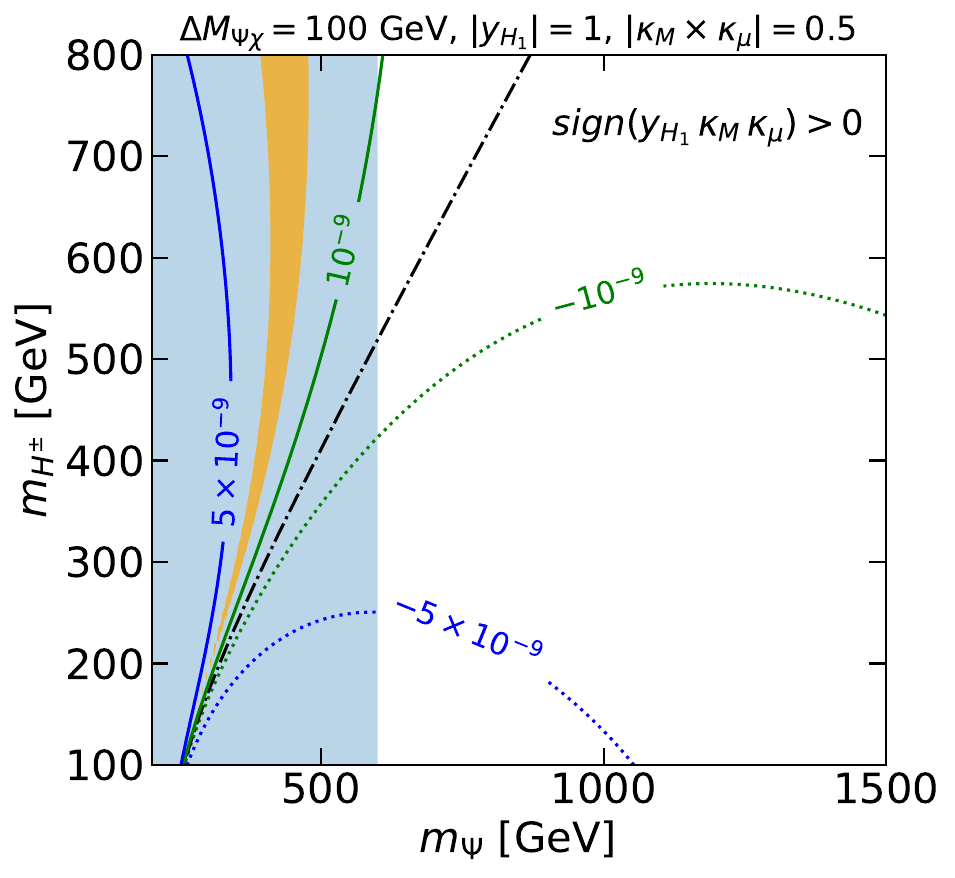}
	\caption{\label{fig:amu} The contribution to $\Delta a_{\mu}$ in this model on the $(m_\Psi, m_{H^\pm})$ plane. We fixed $\Delta M_{\Psi\chi} = 100$ GeV, $|y_{H_1}| = 1$ and $|\kappa_M \times \kappa_\mu| = 0.5$. The sign of $y_{H_1}\kappa_M\kappa_\mu$ is fixed to be negative in the left panel while it is positive in the right panel. The solid blue, green, dotted blue and green represent the values of $\Delta a_{\mu}$ of $5\times 10^{-9}$, $10^{-9}$, $-10^{-9}$ and $-5\times 10^{-9}$, respectively. The dash-dotted black line indicates $\Delta a_{\mu} = 0$, where a complete cancellation between the two terms in the brackets of Eq.~\ref{eq:deltamu} occurs. The orange band is the $1\sigma$ region favored to the discrepancy between the current experiment measurements and the SM results where the HVP is calculated using the data-driven method~\cite{Aoyama:2020ynm}. The light blue region is excluded by the LHC constraints. 
 }
\end{figure}

The dominant contribution to $\Delta a_{\mu}$ in this model arises from the one-loop chirally enhanced Feynman diagram depicted in Fig.~\ref{fig:mu_g2}. This contribution can be expressed in the form~\cite{Calibbi:2018rzv,Arcadi:2021cwg}:
\begin{equation}
\label{eq:deltamu}
\Delta a_{\mu} \simeq -\frac{\kappa_M \kappa_{\mu} m_{\mu}}{4\pi^2 m^2_{H^{\pm}}} \sin 2\alpha \left[ m_{n_1} \tilde{H}(x_1) - m_{n_2} \tilde{H}(x_2) \right],
\end{equation}
where $x_i = \left( m_{n_i}/m_{H^{\pm}}\right)^2$ and $\tilde{H}(x) = \left( x^2 - 1 - 2x\text{log}x \right) / \left[ 8(x-1)^3 \right]$. Due to the effect of chiral enhancement, $\Delta a_{\mu}$ is proportionate to $m_{n_1, n_2}$. 
In Fig.~\ref{fig:amu}, we show contours of $\Delta a_{\mu}$ spanned in the $(m_\Psi, m_{H^\pm})$ plane. Here we fixed values of $\Delta M_{\Psi\chi} = 100$ GeV, $|y_{H_1}| = 1$, and $|\kappa_M \times \kappa_\mu| = 0.5$. 
To achieve a positive and substantial $\Delta a_{\mu}$ from Eq.~(\ref{eq:deltamu}), one possible scenario involves $\kappa_M \times \kappa_{\mu} \times y_{H_1} < 0$ and $m_{H^{\pm}} \ll m_{n_2}$ as shown in the left panel of Fig.~\ref{fig:amu}.  
When $\kappa_M \times \kappa_{\mu} \times y_{H_1} > 0$, the sign of $\Delta a_{\mu}$ flips, as shown in the right panel of Fig.~\ref{fig:amu}. In this case, the region favored by the discrepancy in $\Delta a_{\mu}$ (orange band) is already excluded by the LHC constraints on $\Psi$ mass, i.e $m_{\Psi} > 600$ GeV.

We note that the contributions from other subdominant terms to $\Delta a_{\mu}$ are notably smaller than the one mentioned above, we have omitted their contributions in this study~\cite{Calibbi:2018rzv,Arnan:2019uhr}. On the other hand, the contributions solely from the charged scalar to $\Delta a_{\mu}$ are negative and negligible as the situation in the inert two Higgs doublet model, thus we can disregard them. 

\begin{figure}[h]
\centering 
\includegraphics[width=0.8\textwidth]{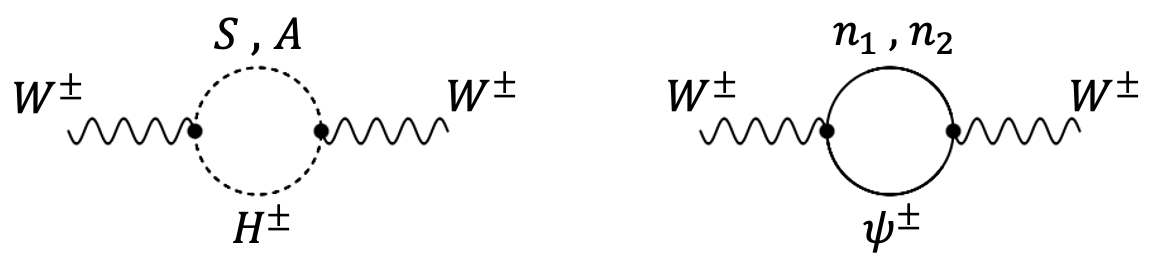}
\caption{The Feynman diagrams for the one-loop $W$ boson mass corrections with the inert Higgs sector and new Dirac fermions.}
\label{fig:feynman_mw}
\end{figure}

Except for the contributions to $\Delta a_{\mu}$, this model can also provide corrections to the squared $W$ boson mass.
In Fig.~\ref{fig:feynman_mw}, we show two representative Feynman diagrams which can contribute to $m_W$ in this model. 
Generally, $\Delta m^2_W$ can be expressed in terms of the oblique parameters $\cal S$, $\cal T$, and $\cal U$~\cite{Peskin:1991sw} as
\begin{equation}
\Delta m^2_W = \frac{\alpha_{\text{EM}}c^2_W m^2_Z}{c^2_W - s^2_W}\left[ -\frac{\cal S}{2} + c^2_W {\cal T} + \frac{c^2_W - s^2_W}{4s^2_W}{\cal U}\right],
\end{equation}
where $c_W$ ($s_W$) is the cosine (sine) of the weak mixing angle, the fine structure constant is denoted as $\alpha_{\text{EM}}$, and the $Z$ boson mass is represented as $m_Z$. Since the contributions of new physics to the $\cal U$ parameter are often higher-order corrections and much more suppressed compared to the contributions from the $\cal S$ and $\cal T$ parameters, for the sake of simplicity, we typically ignore the contributions from the $\cal U$ parameter.

In addition to the higher-order contributions from the SM to $\Delta m^2_W$, there are two extra contributions to $\Delta m^2_W$ in this model. The first contributions to $\Delta m^2_W$ in this model come from the inert Higgs doublet sector, and the contributions to $\cal S$ and $\cal T$ can be represented as~\cite{Barbieri:2006dq,Arhrib:2012ia,Swiezewska:2012ej}:
\begin{equation}
{\cal S} = \frac{1}{2\pi}\left[ \frac{1}{6}\ln\left(\frac{m^2_S}{m^2_{H^{\pm}}}\right)-\frac{5}{36}+\frac{m^2_S m^2_A}{3(m^2_A -m^2_S)^2} +\frac{m^4_A (m^2_A -3m^2_S)}{6(m^2_A -m^2_S)^3}\ln\left( \frac{m^2_A}{m^2_S}\right)\right],
\end{equation}
and
\begin{equation}
{\cal T} = \frac{1}{32\pi^2\alpha_{\text{EM}}v^2}\left[ F(m_h, m_A) + F(m_h, m_S) - F(m_A, m_S)\right],
\end{equation}
where
\begin{equation}
F(M_a, M_b) = \frac{M^2_a + M^2_b}{2} - \frac{M^2_a M^2_b}{M^2_a - M^2_b}\ln\frac{M^2_a}{M^2_b}.
\end{equation}
We use the \texttt{2HDMC}~\cite{Eriksson:2009ws} to calculate the $\cal S$ and $\cal T$ parameters from the inert Higgs doublet sector. We observe that if $m_{H^{\pm}} < m_A$, a positive value of $\cal S$ is generated, resulting in reduced corrections to $\Delta m^2_W$. Consequently, there is a preference for the charged Higgs boson to be the heaviest within the inert Higgs doublet sector when a larger $m_W$ is needed. On the other hand, the generation of a positive $\cal T$ is favored with a large $\Delta^{\pm}$ and/or $\Delta^{\pm}-\Delta^0$, leading to an increase in $\Delta m^2_W$.

\begin{figure}[tb]
    % \centering
	\includegraphics[width=0.45\textwidth]{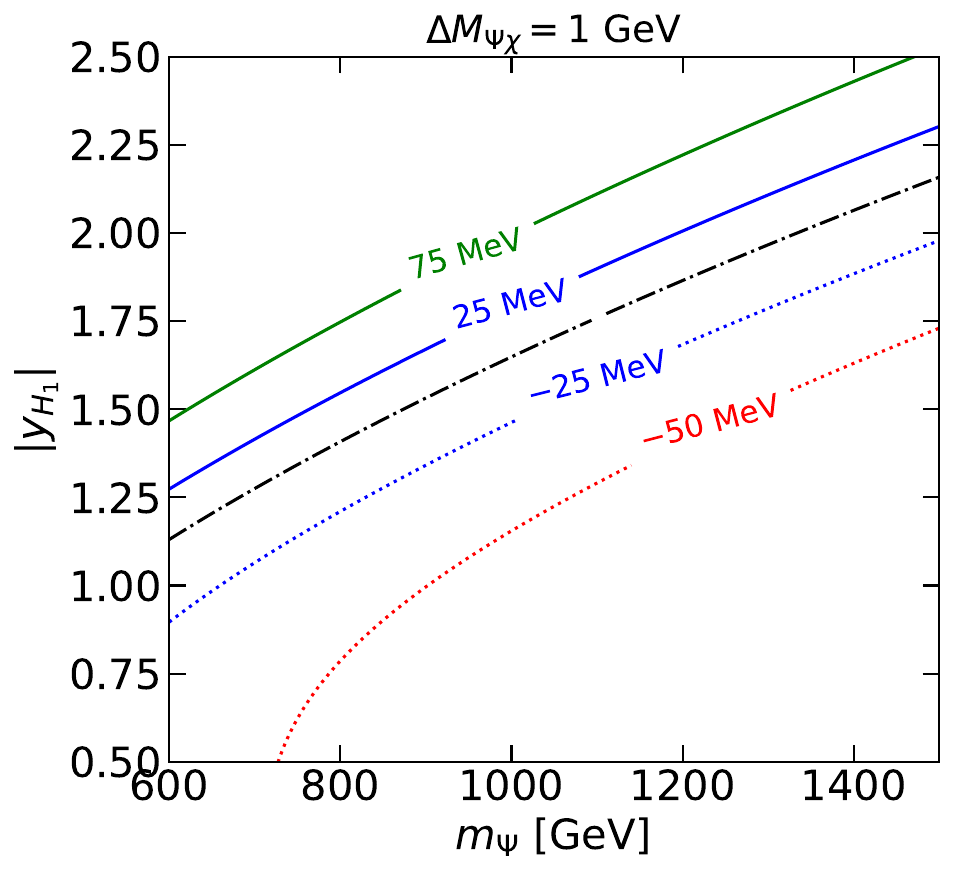}
    \includegraphics[width=0.45\textwidth]{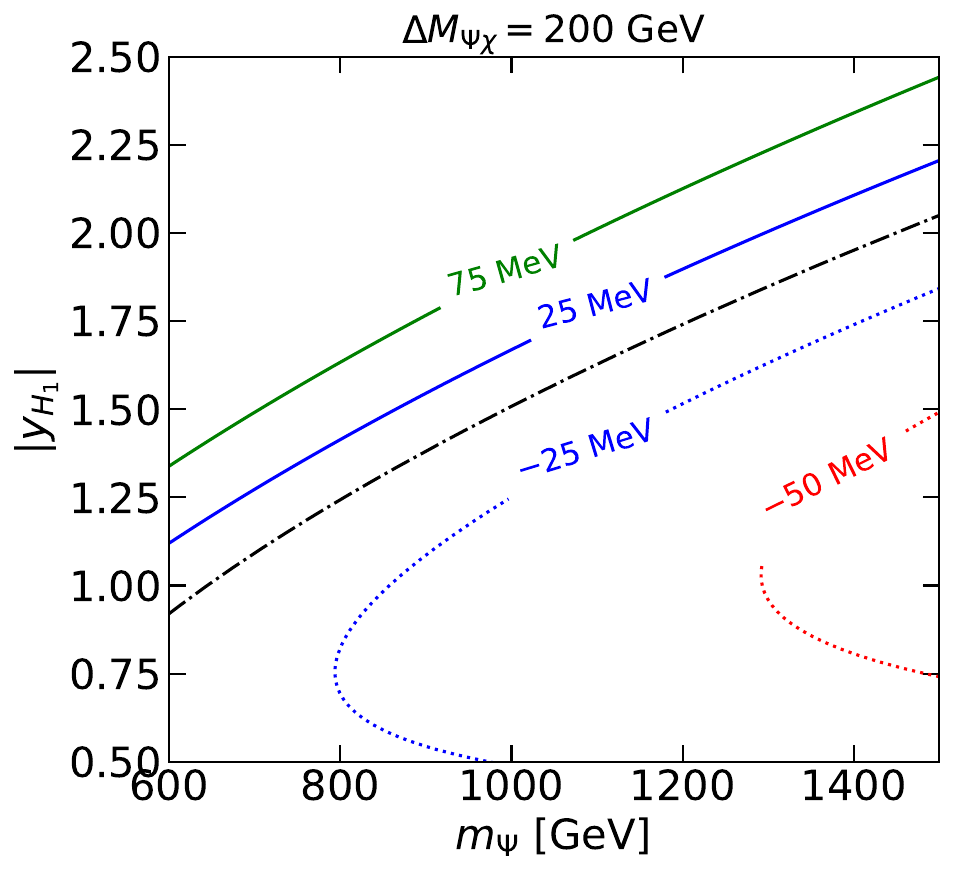}
	\caption{\label{fig:deltaM} Contribution to $\Delta m_W$ from the $\Psi$ field projected onto the $(m_\Psi, |y_{H_1}|)$ plane. In the left panel, we fixed $\Delta M_{\Psi\chi} = 1$ GeV, while in the right panel, $\Delta M_{\Psi\chi} = 200$ GeV. The dotted red, dotted blue, solid blue and solid green  lines represent the values of $\Delta m_{W}$ equal to $-50$ MeV, $-25$ MeV, $25$ MeV, and $75$ MeV, respectively. The dash-dotted black line indicates $\Delta m_{W} = 0.0$. }
\end{figure}

The second contributions to $\Delta m^2_W$ in this model come from the $\Psi$ field, and these additional contributions to the $\cal S$ and $\cal T$ parameters can be summarized as follows~\cite{DEramo:2007anh,Bhattacharya:2018fus}: 
\begin{eqnarray}
{\cal S} &=& \frac{g^2 s^2_W c^2_W}{4\pi^2\alpha_{\text{EM}}} \left[ \frac{1}{3} \left\{ \ln \left(\frac{\Lambda_{\text{EW}}^2}{m_{\psi^{\pm}}^2} \right) -\cos^4 \alpha \ln \left(\frac{\Lambda_{\text{EW}}^2}{m_{n_2}^2} \right)
- \sin^4 \alpha \ln \left(\frac{\Lambda_{\text{EW}}^2}{m_{n_1}^2} \right) \right\} - \frac{1}{2}\sin^2 2\alpha \left\{   
\ln \left(\frac{\Lambda_{\text{EW}}^2}{m_{n_1} m_{n_2}} \right) \right.\right.\nonumber\\ 
 && \left. \left. + \frac{m_{n_1}^4-8 m_{n_1}^2 m_{n_2}^2 +m_{n_2}^4}{9 (m_{n_1}^2-m_{n_2}^2)^2} + \frac{(m_{n_1}^2+m_{n_2}^2)(m_{n_1}^4-4m_{n_1}^2m_{n_2}^2+m_{n_2}^4)}{6(m_{n_1}^2-m_{n_2}^2)^3} 
\ln \left(\frac{m_{n_2}^2}{m_{n_1}^2} \right) \right.\right.\nonumber\\
&& \left. \left. +   \frac{m_{n_1} m_{n_2}(m_{n_1}^2 + m_{n_2}^2)}{2(m_{n_1}^2-m_{n_2}^2)^2} + \frac{m_{n_1}^3m_{n_2}^3}{(m_{n_1}^2-m_{n_2}^2)^3} \ln \left(\frac{m_{n_2}^2}{m_{n_1}^2} \right)   \right\} \right],
\end{eqnarray} 
and
\begin{equation}
{\cal T}=\frac{g^2}{16 \pi^2\alpha_{\text{EM}} M_W^2}\left[ \frac{1}{2}\sin^2 2\alpha ~\Pi^{\prime}(m_{n_1},m_{n_2},0)-2\cos^2\alpha ~\Pi^{\prime}(m_{\psi^{\pm}},m_{n_2},0)-2\sin^2\alpha ~\Pi^{\prime}(m_{\psi^{\pm}},m_{n_1},0)\right]\,,
\end{equation}
 where $\Pi^{\prime}(M_a,M_b,0)$ is given by 
\begin{eqnarray}
\Pi^{\prime} (M_a,M_b,0) &=& -\frac{1}{2}(M_a^2+M_b^2) \ln \left(\frac{\Lambda_{\text{EW}}^2}{M_a M_b} \right) -\frac{1}{4} (M_a^2+ M_b^2)-\frac{(M_a^4+M_b^4)}
{4(M_a^2-M_b^2)} \ln \frac{M_b^2}{M_a^2} \nonumber\\
&& + M_a M_b \left\{ \ln \left(\frac{\Lambda_{\text{EW}}^2}{M_a M_b} \right)+1+ \frac{(M_a^2 +M_b^2)}{2(M_a^2-M_b^2)} \ln  \frac{M_b^2}{M_a^2}  \right\} \,. 
\end{eqnarray}
Here $\Lambda_{\text{EW}}$ is at the electroweak scale and we set $\Lambda_{\text{EW}} = 246$ GeV.

To investigate the influence of the model parameters $m_{\Psi}$, $y_{H_1}$, and $\Delta M_{\Psi\chi}$ on $\Delta m_W$, we analyze the impact of the $\Psi$ field on $\Delta m_W$ in the $(m_\Psi, |y_{H_1}|)$ plane. The results are presented in Fig.~\ref{fig:deltaM} for two cases: $\Delta M_{\Psi\chi} = 1$ GeV (left panel) and $\Delta M_{\Psi\chi} = 200$ GeV (right panel). 
The presence of $\Psi$ field can significantly affect the mass of the $W$ boson. In particular,
the $W$ boson mass increases as $|y_{H_1}|$ becomes larger. Additionally, smaller masses of $\Psi$ and $\chi$, result in a further increase in the $W$ boson mass.  In certain specific regions, the effect from the $\Psi$ field can even dominate the contributions from the inert Higgs doublet sector. 
For instance, at $\Delta M_{\Psi \chi} = 1$ GeV, $m_\Psi = 1$ TeV and $|y_{H_1}| = 2$, the resulting $\Delta m_W \sim 75$ MeV, consistent with the CDF II measurement. Nevertheless, it is important to note that in certain areas of the parameter space, the contribution to $\Delta m_W$ diminishes, as depicted by the dash-dotted black line in Fig.~\ref{fig:deltaM}.

\section{Results  and discussions}
\label{sec:result}

\subsection{Methodology}

%******************************************************************
\begin{table}[t!]
    \centering
    \begin{tabular}{|l|l|}
    \hline\hline
        Likelihood type & Constraints \\
        \hline
        Step &  perturbativity, stability, unitarity~\cite{Eriksson:2009ws}\\
        Step & LEP-II~\cite{Agashe:2014kda}, OPAL~\cite{Abbiendi:2003ji}\\
        Half-Gaussian & current-LZ~\cite{LUX-ZEPLIN:2022xrq}\\
        Half-Gaussian & Higgs invisible decays~\cite{ATLAS:2023tkt}\\ 
        Half-Gaussian & Higgs exotic decays~\cite{ATLAS:2020qdt}\\
        Gaussian & relic abundance~\cite{Aghanim:2018eyx}\\
        Gaussian & $R_{\gamma\gamma}$~\cite{ATLAS:2018doi}\\
        \hline
        Gaussian & 
        $m_{W,{\rm CDF\,II}}$~\cite{CDF:2022hxs} or $m_{W,{\rm PDG}}$~\cite{ParticleDataGroup:2022pth} \\ 
        \hline
        Gaussian & 
        $\Delta a_{\mu}^{e^+e^-\text{ HVP}}$~\cite{Aoyama:2020ynm} or $\Delta a_{\mu}^{\text{ Lattice HVP}}$~\cite{Borsanyi:2020mff} \\   
        \hline\hline
    \end{tabular}
    \caption{Likelihood distributions and constraints used in our analysis.}
    \label{tab:likelihood}
\end{table}
%*******************************************************************
We employ the ``\textit{Profile Likelihood}'' method~\cite{Rolke:2004mj} 
to eliminate nuisance parameters and illustrate the two dimensional contours. The specific experimental constraints are listed
in Table~\ref{tab:likelihood}. First of all, we use the theoretical bounds as well as LEP-II and OPAL limits as the hard cuts. Secondly, for Gaussian functions, we employ the following chi-square equation:
\begin{equation}
    \chi^2 = \left( \frac{\mu - \mu_\mathrm{exp}}{\sigma} \right)^2~{\rm and}~
    \sigma = \sqrt{\sigma^2_\mathrm{theo} + \sigma^2_\mathrm{exp}},
\end{equation}
where $\mu$ represents the theoretical prediction and $\mu_\mathrm{exp}$ denotes the experimental central value. However, for Half-Gaussian functions where a null signal is expected,  we set $\mu_\mathrm{exp}=0$.
The uncertainty $\sigma$ incorporates both theoretical and experimental errors. By summing the individual  $\chi^2$ of these constraints, we obtain the total $\chi^2_\mathrm{tot}$ as 
\begin{equation}
    \chi^2_\mathrm{tot} = \chi^2_\mathrm{a_{\mu}} + \chi^2_\mathrm{m_{w}} + \chi^2_\mathrm{\Omega h^{2}} + \chi^2_\mathrm{h_{exo}} +\chi^2_\mathrm{h_{inv}} + \chi^2_\mathrm{R_{\gamma\gamma}} + \chi^2_\mathrm{LZ}, 
\end{equation} 
where $\chi^2_\mathrm{a_{\mu}}$, $\chi^2_\mathrm{m_{w}}$, $\chi^2_\mathrm{\Omega h^{2}}$, $\chi^2_\mathrm{h_{exo}}$, $\chi^2_\mathrm{h_{inv}}$, $ \chi^2_\mathrm{R_{\gamma\gamma}}$ and $\chi^2_\mathrm{LZ}$ represent the $\chi^2$ from muon $g-2$, $W$-boson mass, DM relic density, Higgs boson exotic decays, Higgs boson invisible decays, signal strength of Higgs boson to diphoton and DM direct detection, respectively.

To elucidate the effects of quality anomalies from muon $g-2$ and $W$-boson mass, we conduct four sets of numerical scans: 
\begin{enumerate}
\item $m_{W, \text{CDF II}}=80.4335\pm 0.0094$ GeV and $\Delta a_{\mu}^{e^+e^-\text{ HVP}} = 249(48)\times 10^{-11}$; 
\item $m_{W,\text{PDG}} = 80.377\pm 0.012$ GeV and $\Delta a_{\mu}^{e^+e^-\text{ HVP}} = 249(48)\times 10^{-11}$; 
\item $m_{W, \text{CDF II}}=80.4335\pm 0.0094$ GeV and $\Delta a_{\mu}^{\text{Lattice HVP}} = 105(59)\times 10^{-11}$; 
\item $m_{W,\text{PDG}} = 80.377\pm 0.012$ GeV and $\Delta a_{\mu}^{\text{Lattice HVP}} = 105(59)\times 10^{-11}$.
\end{enumerate} 
Here $m_{W, \text{CDF II}}$ and $m_{W, \text{PDG}}$ are the $W$ boson mass results from the CDF II experiments and from the Particle Data Group~\cite{ParticleDataGroup:2022pth}, respectively. The muon $g-2$ results, $\Delta a_{\mu}^{e^+e^-\text{ HVP}}$ and $\Delta a_{\mu}^{\text{Lattice HVP}}$, correspond to the HVP contributions obtained via data-driven and lattice calculation methods, respectively.

On the other hand, we also compare our allowed parameter space with several DM indirect detection constraints. These include the bounds established by the Fermi LAT observations of dwarf Spheroidal galaxies~\cite{Fermi-LAT:2016uux}, the signal regions of the Fermi LAT Galactic Center $\gamma$-ray excess~\cite{Hooper:2010mq,Zhou:2014lva,Calore:2014xka,Daylan:2014rsa}, and the Alpha Magnetic Spectrometer AMS-02 experiment antiproton excess~\cite{Cui:2016ppb,Cuoco:2016eej,Cui:2018klo,Cholis:2019ejx}.

We conduct Markov Chain Monte Carlo (MCMC) scans using the \texttt{EMCEE}~\cite{ForemanMackey:2012ig} code and obtain around  $\mathcal{O}(4.5\times 10^6)$ data points. The confidence intervals are derived from the tabulated values of $\Delta\chi^2\equiv -2\ln(\mathcal{L/L}_{\rm{max}})$ where $\mathcal{L}_{\rm{max}}$ is the likelihood at the best-fits. For a two-dimensional plot, the 95\% confidence level (C.L.) region ($2\sigma$ allowed region) is defined as $\Delta\chi^2 \leq 5.99$, assuming an approximate Gaussian likelihood.

\subsection{Results}
\label{sec:results}

\begin{figure}[htp]
  \centering
  \includegraphics[width=8.1cm,height=8.1cm]{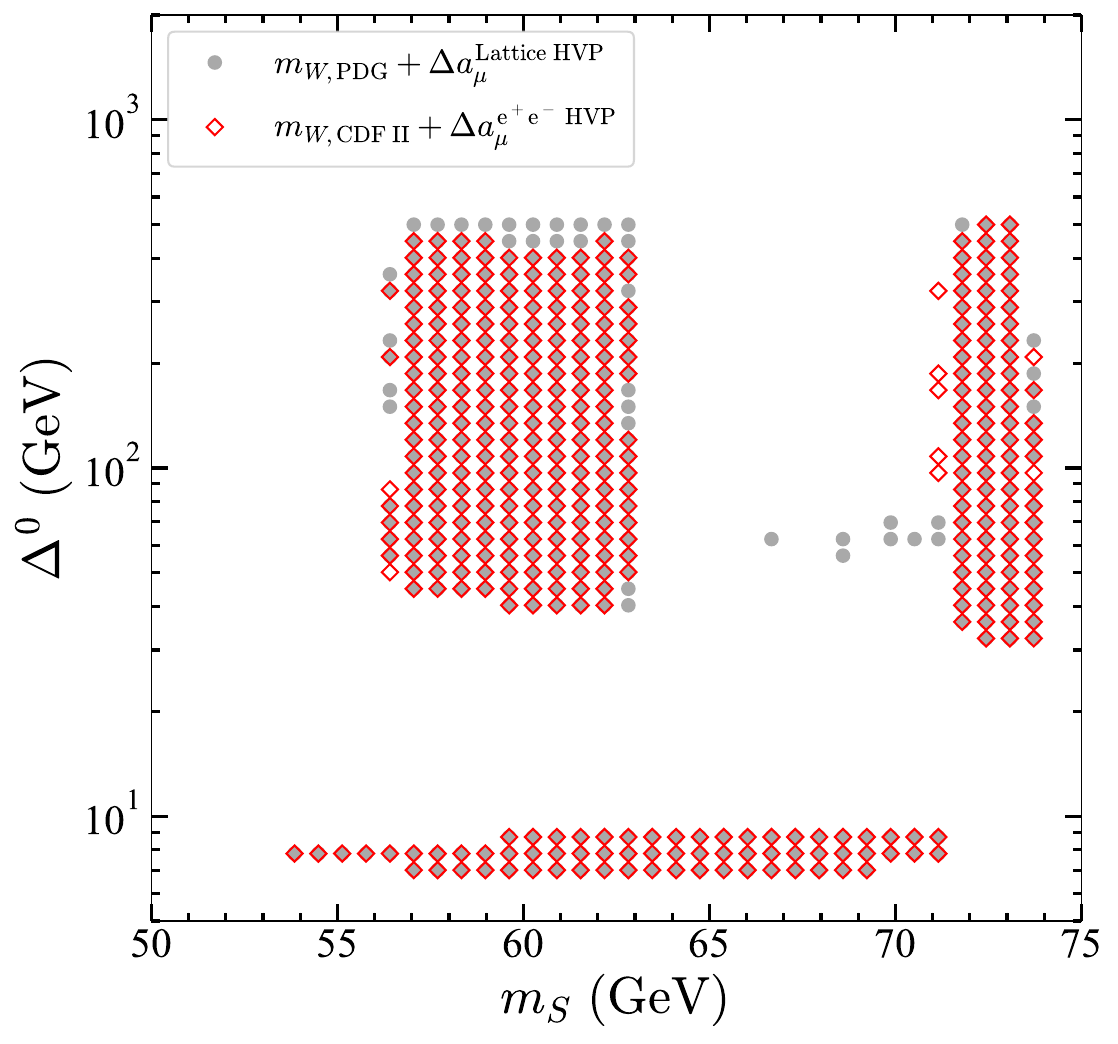}
  \includegraphics[width=8.1cm,height=8.1cm]{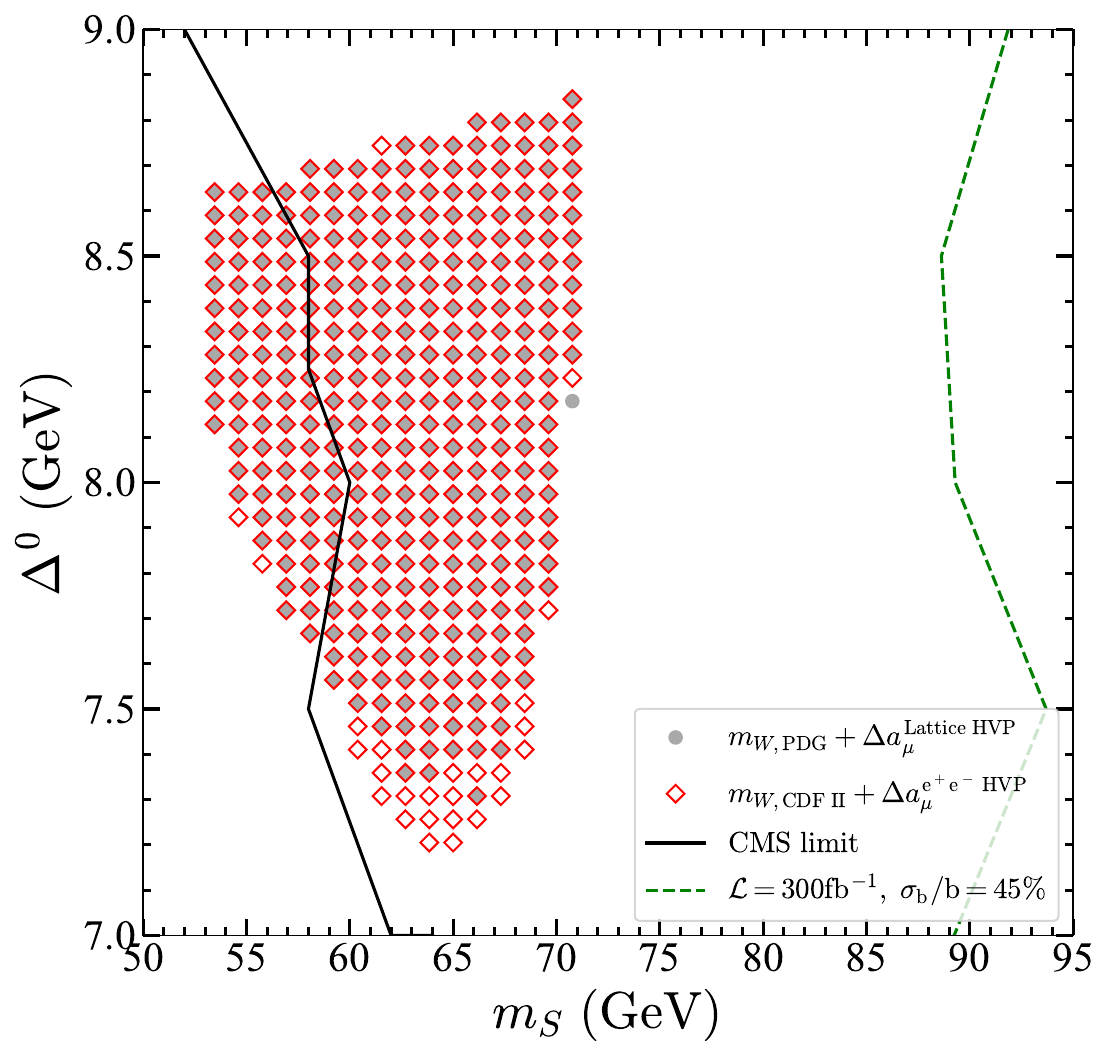}
  \includegraphics[width=8.1cm,height=8.1cm]{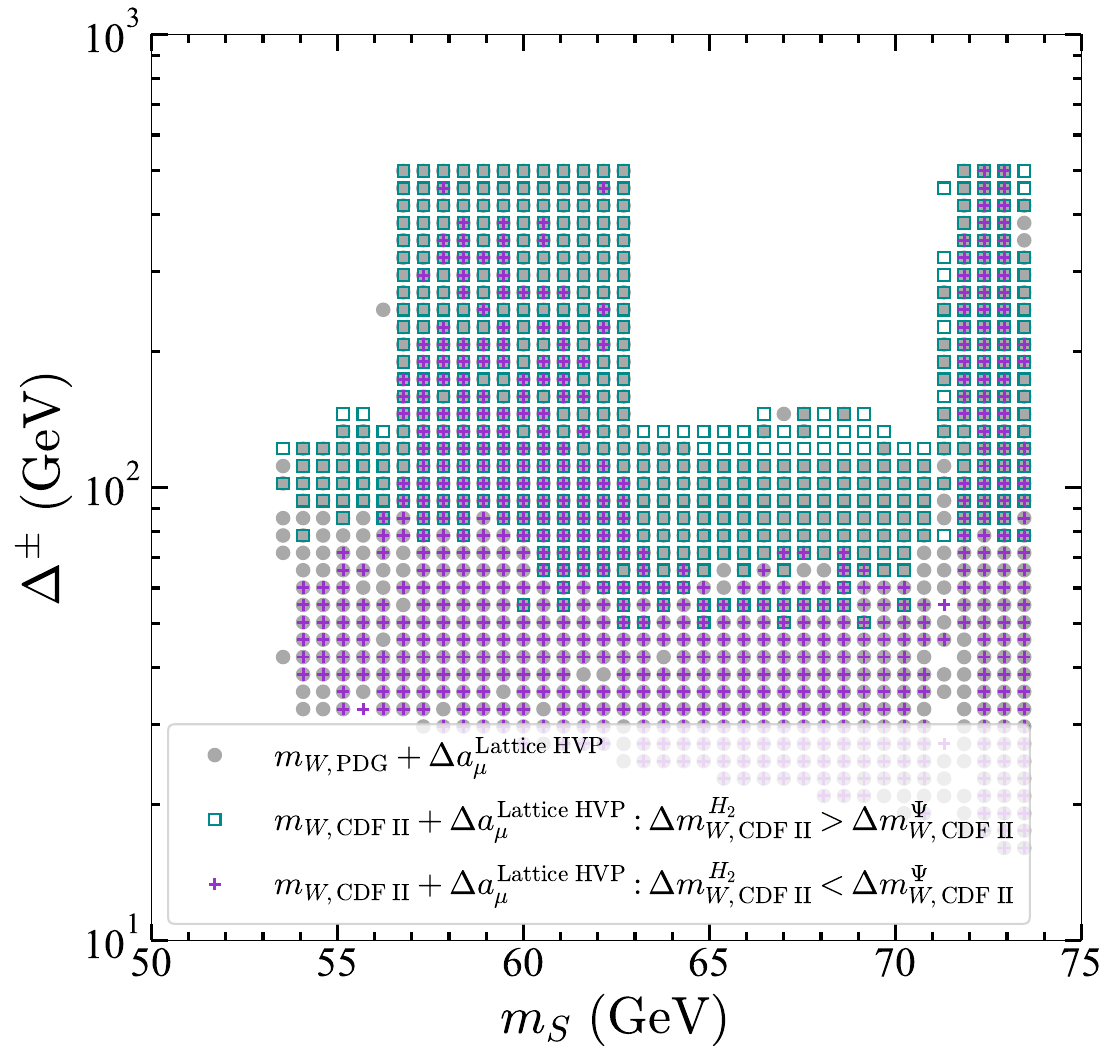}
  \includegraphics[width=8.1cm,height=8.1cm]{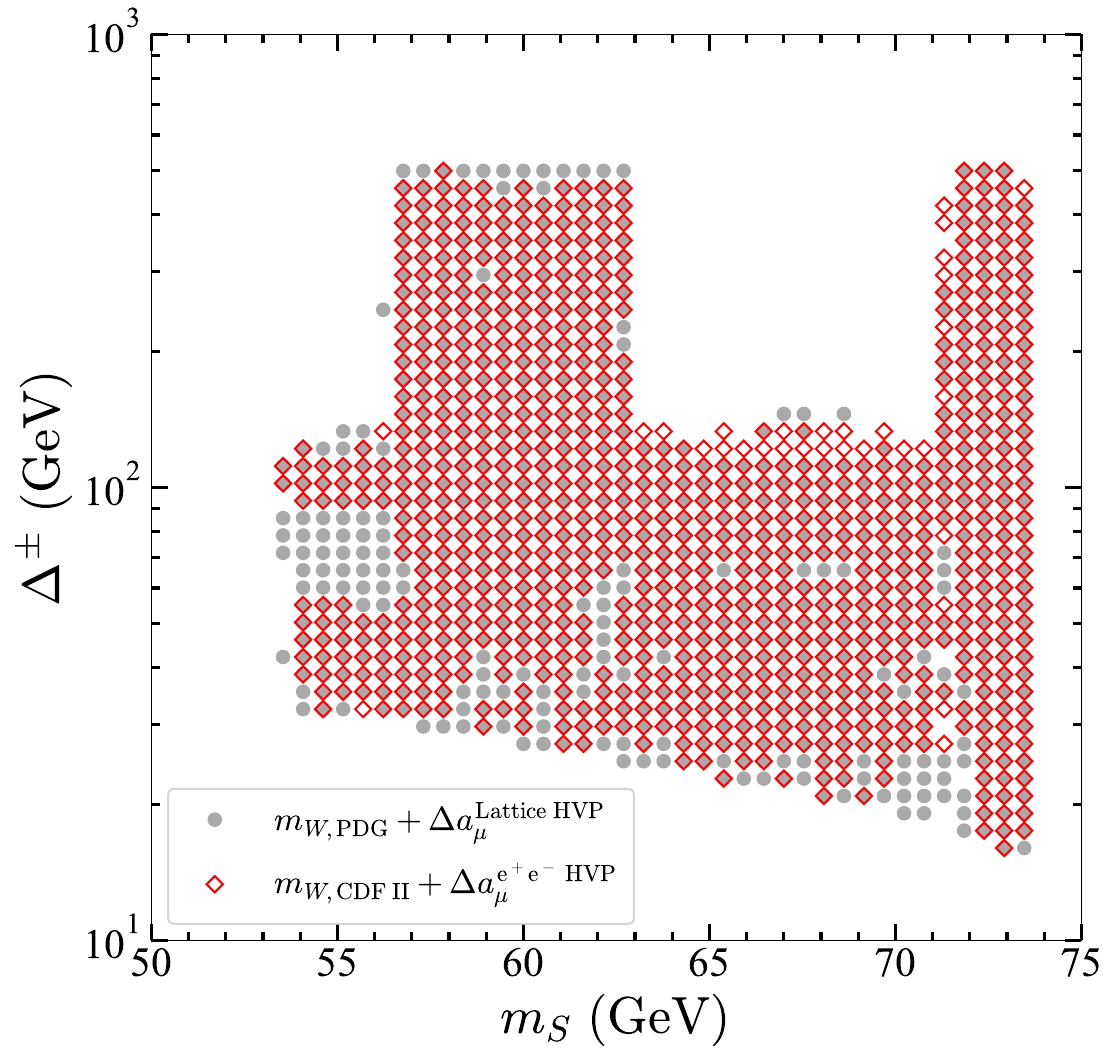}
  \caption{The 95$\%$ C.L. regions on the panels of $m_s$  versus $\Delta^0$ (upper panels) and $\Delta^\pm$ (lower panels).
  The gray region favored by the data of $m_{W,\text{PDG}}$ and $\Delta a_\mu^\text{Lattice HVP}$.
  The red diamond region represents the combined results of $m_{W,\text{CDF II}}$ and $\Delta a_{\mu}^{e^+e^-\text{ HVP}}$.
  The green box indicates the region where the contribution to $m_{W,\text{CDF II}}$ satisfies $\Delta m^{H_2}_{W,\text{CDF II}} > \Delta m^\Psi_{W,\text{CDF II}}$, while the purple cross represents the region where $\Delta m^{H_2}_{W,\text{CDF II}} < \Delta m^\Psi_{W,\text{CDF II}}$.
 The solid black and dashed green lines in the upper right panel indicate the current limit from CMS and future probe from LHC Run 3, respectively.
  \label{Fig:ms_mH} 
  }
\end{figure}

In this section, we present the $95\%$ C.L. regions for the total likelihoods, as shown in Table~\ref{tab:likelihood}. 
Unless otherwise specified, the gray circles and red hollow diamonds represent scenarios where the model parameter regions favor $m_{W,\text{PDG}}\,+\,\Delta a_\mu^\text{Lattice HVP}$ and $m_{W,\text{CDF II}}\,+\,\Delta a_{\mu}^{e^+e^-\text{HVP}}$, respectively.

Within the range of model parameters we consider, the DM mass $m_S$ is found to lie between 54 and $74\,\text{GeV}$, irrespective of whether the anomalies in the muon magnetic moment and the $W$ boson mass are taken into account.
In the upper left panel of Fig.~\ref{Fig:ms_mH}, we show the $95\%$ C.L. region for $m_S$ and $\Delta^0$.
The DM mass clearly falls into three distinct regions, each corresponding to a different dominant DM annihilation mechanism in the early universe. 
For $m_S \approx m_h/2$, this corresponds to the Higgs resonance.
The $\Delta^0\lesssim 10\,\text{GeV}$ indicates $SA$ co-annihilation, while $ 71 \lesssim m_S \lesssim 74 \,\text{GeV}$ corresponds to the $SS \to WW^*$ process.
The gray region contains a few scattered points, which arise from DM annihilation processes dominated by neutrino final states via the exchange of $n_{1,2}$. However, the significant overlap between the gray and red regions suggests that the anomalies in the muon magnetic moment and the $W$ boson mass have only a slight effect on the dominant DM annihilation mechanisms.

For the $SA$ co-annihilation mechanism, we present an enlarged view of the $95\%$ C.L. regions for $m_S$ versus $\Delta^0$ in the upper right panel of Fig.~\ref{Fig:ms_mH}.
The solid black line represents the exclusion limits from the CMS search for two or three soft leptons at the $13\,\text{TeV}$ LHC, with an integrated luminosity of $137\,\text{fb}^{-1}$~\cite{CMS:2021edw}.
The green dashed line illustrates the projected $2\sigma$ sensitivity, assuming a conservative background uncertainty of $\sigma_b / b = 45\%$, for an integrated luminosity of $300\,\text{fb}^{-1}$ in one of the signal regions.
With the forthcoming LHC luminosity upgrades, the entire $SA$ coannihilation region is expected to be thoroughly tested. Detailed calculations and discussions regarding collider experiments can be found in the latter part of this section and in Sec.~\ref{sec:future}.

After imposing all constraints, $\Delta^\pm$ is limited to the range of $17-500\,\text{GeV}$ for the $m_{W,\text{PDG}}+\Delta a_\mu^\text{Lattice HVP}$ scenario, as shown by the gray regions in the lower panels of Fig.~\ref{Fig:ms_mH}.
In this model, the extra corrections to $W$ boson mass arise from loop contributions involving the inert Higgs fields $H_2$ and new fermion fields $\Psi$ (as depicted in  Fig.~\ref{fig:feynman_mw}).
If the $H_2$ sector serves as the dominant contribution, an larger mass splitting between the charged Higgs $H^\pm$ and the neutral Higgs $S$ is expected.
In a more conservative scenario, we require the $\Delta m^{H_2}_{W,\text{CDF II}} > \Delta m^\Psi_{W,\text{CDF II}}$, which corresponds to the green box regions in the lower left panel of Fig.~\ref{Fig:ms_mH}.
However, if $m_{W,\text{CDF II}}$ is primarily influenced by corrections from the $\Psi$ fields, the heavy charge Higgs $H^\pm$ becomes unnecessary. 
In this case, we impose $\Delta m^{H_2}_{W,\text{CDF II}} < \Delta m^\Psi_{W,\text{CDF II}}$, with the allowed regions denoted by purple crosses. 
Clearly, compared to the scenario dominated by the inert Higgs, the $95\%$ C.L. regions for $\Delta^\pm$ shift downward.
Finally, we present results that jointly consider both $m_{W,\text{CDF II}}$ and $\Delta a_{\mu}^{e^+e^-\text{ HVP}}$, shown as red diamond regions in the lower right panel of Fig.~\ref{Fig:ms_mH}. 
Clearly, the constraint results are not more stringent than the those in the gray regions. This model relaxes the requirement for a larger $\Delta^\pm$ contribution to $m_{W,\text{CDF II}}$ compared to the traditional inert 2HDM.

\begin{figure}[t]
  \centering
  \includegraphics[width=8.1cm,height=8.1cm]{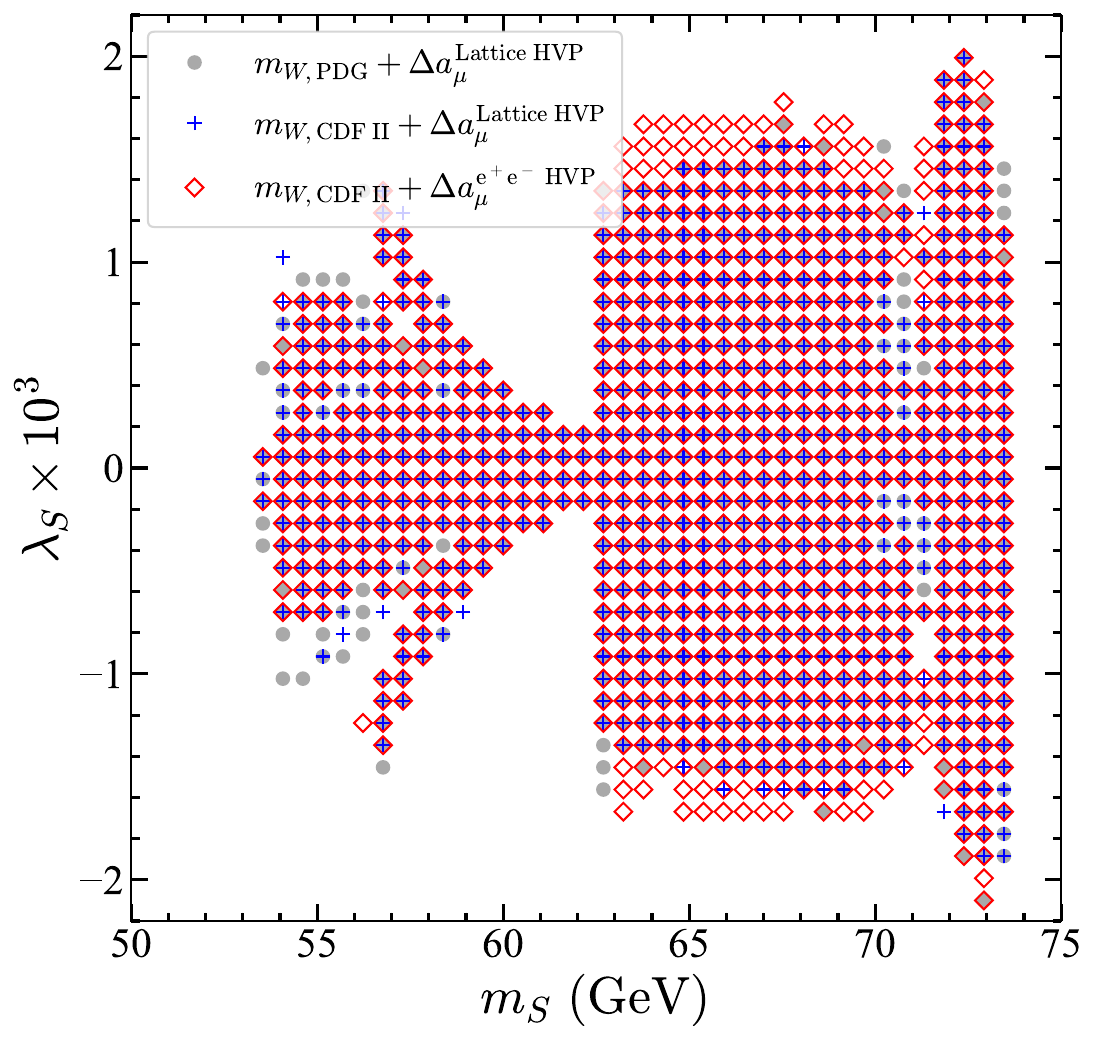}
  \includegraphics[width=8.1cm,height=8.1cm]{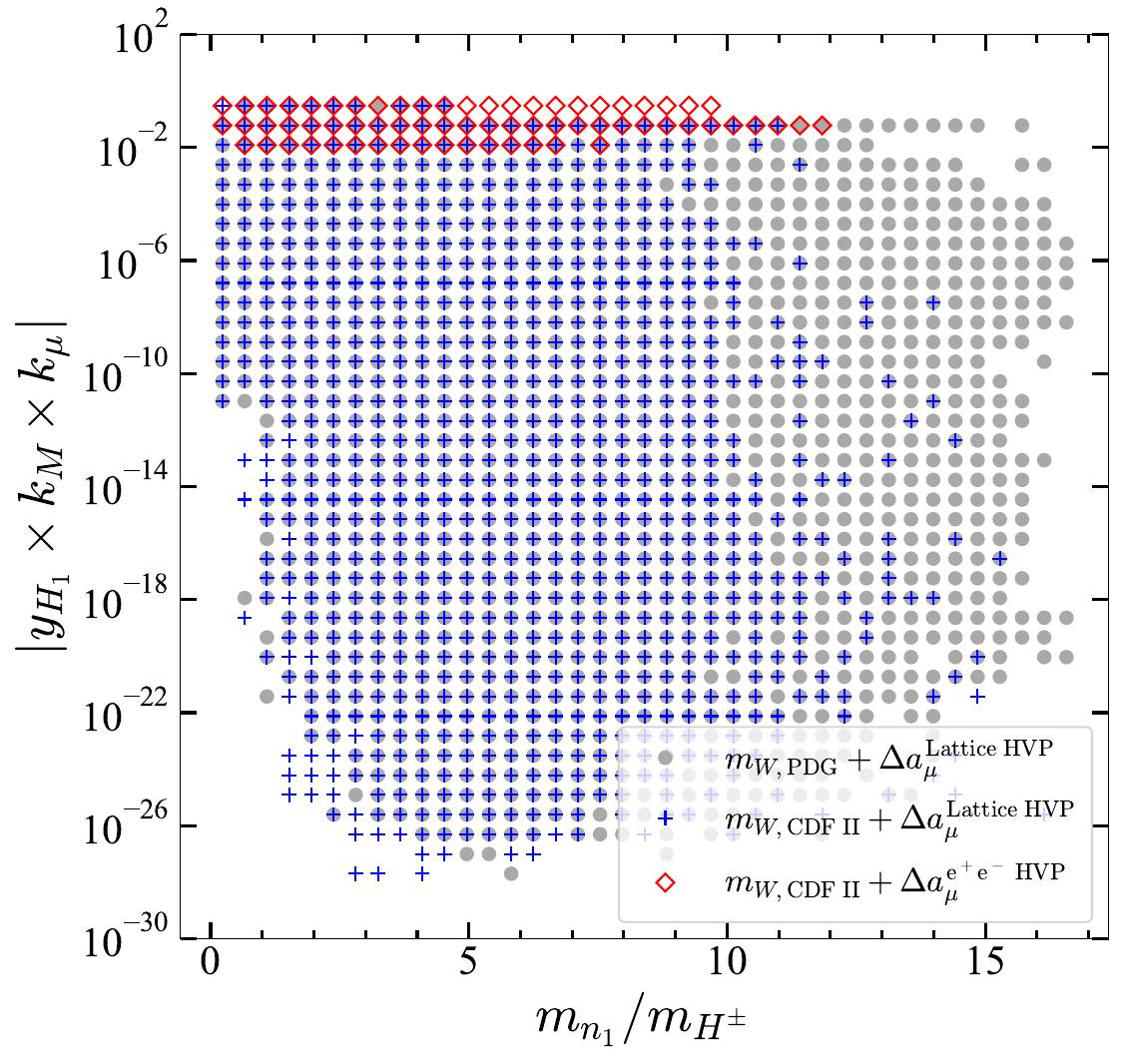}
  \caption{Left panel: The $95\%$ C.L. regions for the Higgs-DM-DM coupling $\lambda_S\times10^3$ as a function of the DM mass $m_S$; Right panel: The correlation  between the product of coupling coefficients $|y_{H_1}\times k_M \times k_\mu|$ and mass spectra $m_{n_1}/m_{H^\pm}$. The blue cross region favor the CDF II $W$ boson mass, while the definitions of other colors are the same as in Fig.~\ref{Fig:ms_mH}.}
  \label{Fig:couplings}
\end{figure}

In the left panel of Fig.~\ref{Fig:couplings}, we present the 95\% C.L. regions for the Higgs-DM-DM coupling $\lambda_S$ as a function of the DM mass. We find that $\lambda_S$ must remain small, mainly due to the Planck relic density constraint, as a larger coupling leads to a shortage in DM abundance. 
As expected, the viable range of $\lambda_S$ is not significantly affected when the $m_{W,\text{PDG}} + \Delta a_{\mu}^\text{Lattice HVP}$ scenario is imposed, as shown by the gray region. 
Notably, around $m_S \simeq 62$ GeV, the coupling $\lambda_S$ exhibits a bottleneck shape, a feature driven by the Higgs resonance mechanism. As the DM mass moves away from the resonance, increasingly larger values of $\lambda_S$ are needed to open up other annihilation channels.

In the same panel, we observe a marginal narrowing of the allowed region when the $m_{W,\text{CDF II}} + \Delta a_{\mu}^\text{Lattice HVP}$ scenario 
is applied, as indicated by the blue regions.
Interestingly, when the $m_{W,\text{CDF II}} + a_{\mu}^{e^+e^-\text{ HVP}}$ scenario is considered, the constraints on $\lambda_S$ are slightly relaxed compared to the other scenarios (red regions). This relaxation occurs because the contribution from $\Delta a_{\mu}^{e^+e^-\text{ HVP}}$ opens DM  annihilation channels dominated by neutrino final states, mediated via the exchange of $n_{1,2}$.

\begin{figure}[htp]
  \centering
  \includegraphics[width=8.1cm,height=8.1cm]{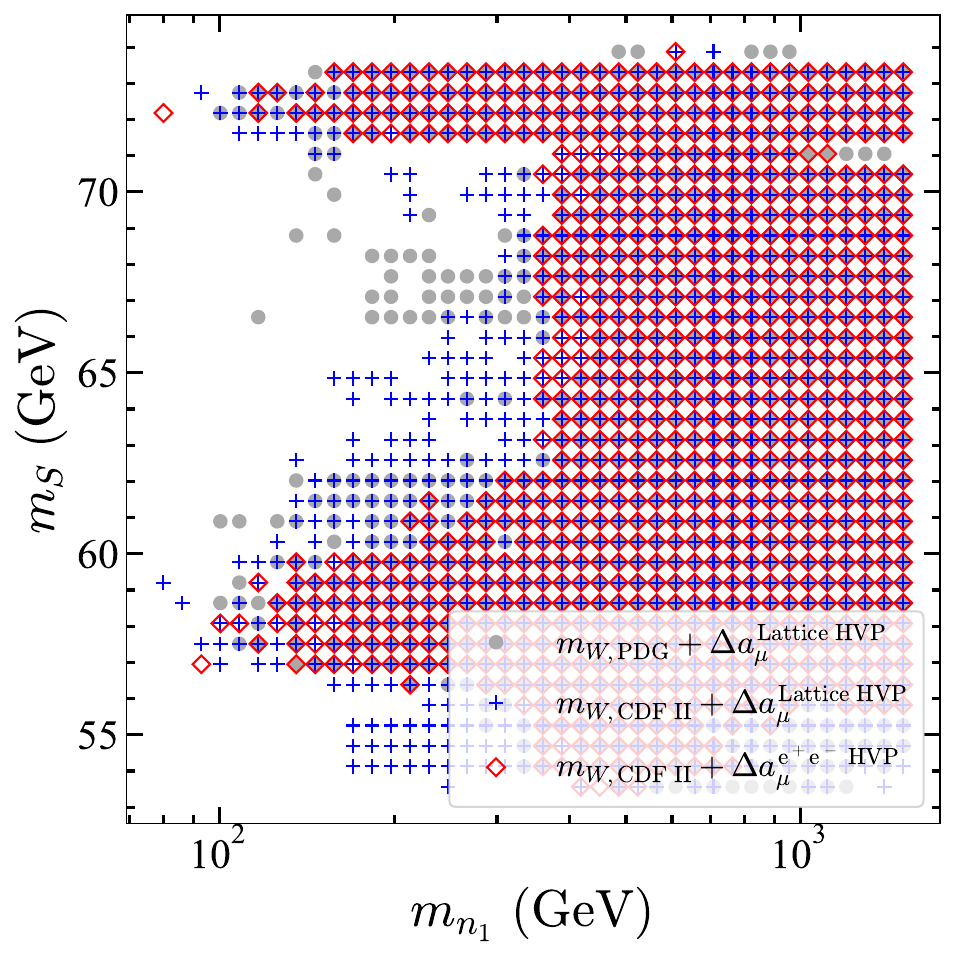}
  \caption{The $95\%$ C.L. regions for the neutral dark particle mass $m_{n_1}$ and the DM mass $m_S$. The color definitions are consistent with those in Fig.~\ref{Fig:couplings}.}
  \label{Fig:mn1}
\end{figure}

On the other hand, we show the allowed regions on the mass ratio $m_{n_1}/m_{H^\pm}$ versus coupling coefficients $|y_{H_1}\times k_M \times k_\mu|$ in the right panel of Fig.~\ref{Fig:couplings}. 
We find that the mass ratio $m_{n_1}/m_{H^\pm} \gtrsim 0.4$ in nearly all of the allowed regions. 
As indicated by Eq.~(\ref{eq:deltamu}), this results in a positive contribution to $\Delta a_\mu$, aligning with the findings presented in the left panel of Fig.~\ref{fig:amu}.
The upper limit on $m_{n_1}/m_{H^\pm} \lesssim 17$ is due to the range of the model parameters under consideration (gray region).
The CDF II $W$ mass requires a heavier $H^\pm$, resulting in a reduced ratio of $m_{n_1}/m_{H^\pm}$ as shown in the blue region. 
Additionally, considering $\Delta a_{\mu}^{e^+e^-\text{ HVP}}$, which deviates from the SM prediction by 5.1$\sigma$, the coupling coefficients must be sizable to enhance contributions to the muon magnetic moment. Consequently, $|y_{H_1}\times k_M \times k_\mu| \gtrsim 10^{-2}$, as indicated by the red regions in the right panel of Fig.~\ref{Fig:couplings}. Moreover, in order to clearly illustrate the mass range of the neutral dark particle $n_1$, we present in Fig.~\ref{Fig:mn1} the mass spectra of $n_1$ and the DM candidate $S$.
The mass of $n_1$ spans a range from $80\,\rm{GeV}$ to $1.5\,\rm{TeV}$.

\begin{figure}[htp]
  \centering
  \includegraphics[width=8.1cm,height=8.1cm]{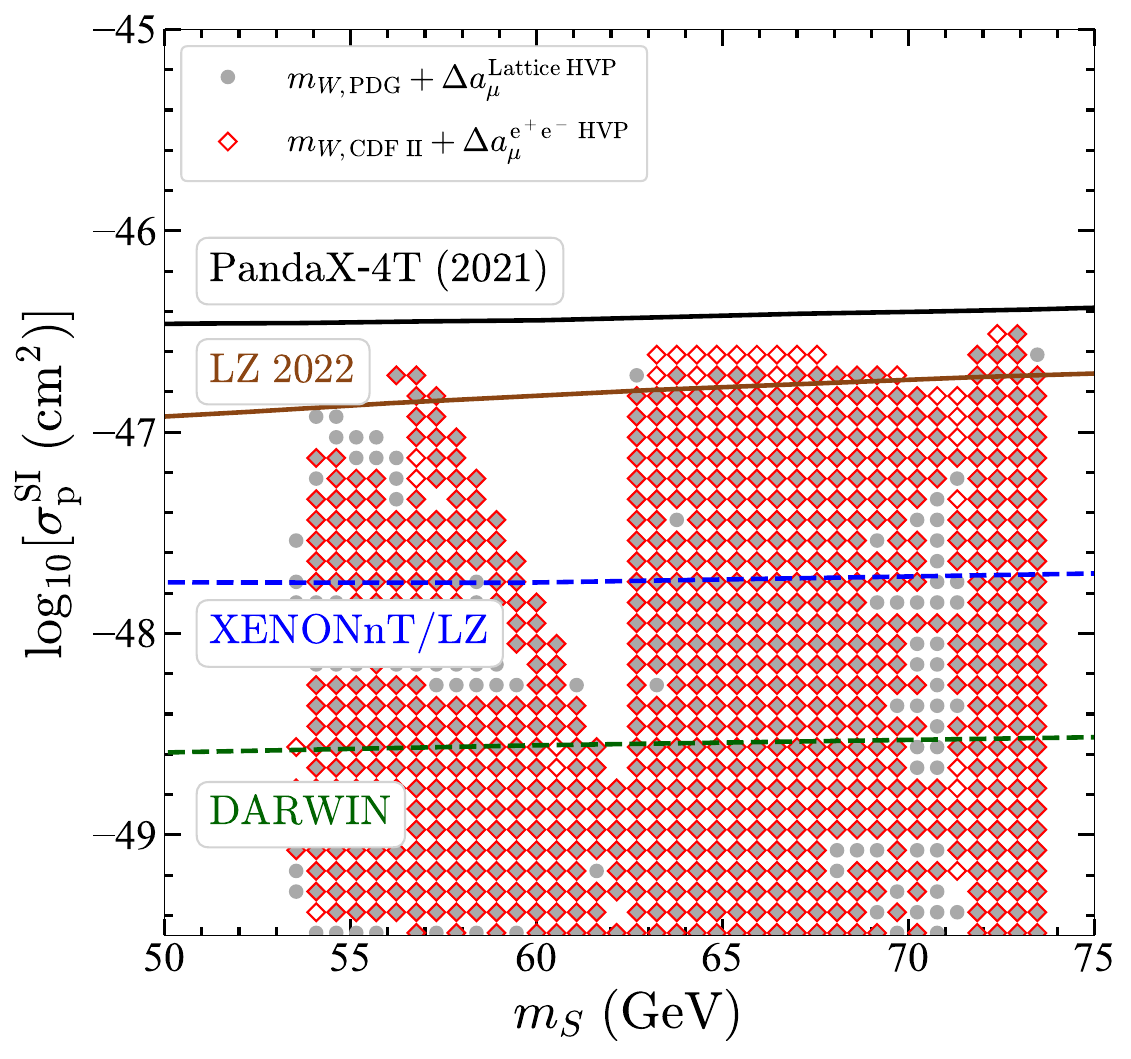}
  \includegraphics[width=8.1cm,height=8.1cm]{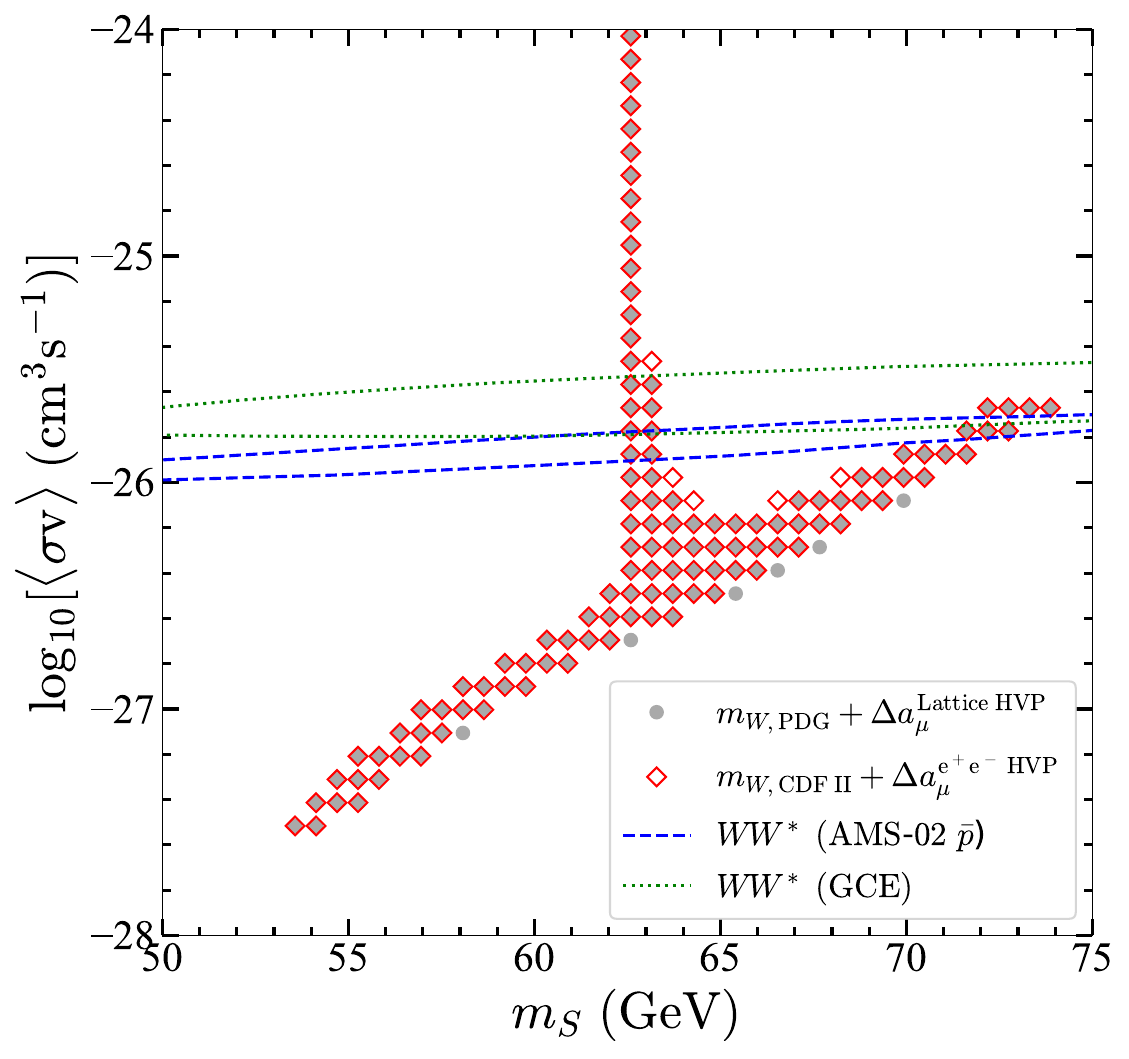}  
  \caption{Left panel: The $95\%$ C.L. regions from all new constraints projected on the ($m_S$, $\sigma_p^{\text{SI}}$) panel. The solid black and brown lines represent the current constraints set by PandaX-4T (2021)~\cite{PandaX-4T:2021bab} and LZ 2022~\cite{LZ:2022ufs}, respectively. The dashed blue and green lines are the expected detection sensitivity of XENONnT/LZ~\cite{XENON:2020kmp,LZ:2015kxe} and DARWIN~\cite{Schumann:2015cpa}. Right panel: The $95\%$ C.L. regions on the ($m_S$, $\langle \sigma v\rangle$) panel. The dotted green and dashed blue contours indicate the parameter regions that can explain the GCE and AMS-02 $\overline{p}$ excesses through the DM annihilation channel $SS\to WW^*$~\cite{Zhu:2022tpr}. }
  \label{Fig:exp2}
\end{figure}

In the left panel of Fig.~\ref{Fig:exp2}, we present the allowed regions on the plane of the DM-proton scattering cross-section and the DM mass. We note that the recent results from LZ experiment~\cite{LZ:2022ufs} (brown solid line) has imposed tighter constraints on the DM-proton scattering cross-section compared to the results in Ref.~\cite{Fan:2022dck}.
In this model, the only process that significantly contributes to the DM direct detection rate at tree level is elastic scattering via t-channel Higgs boson exchange. 
Therefore, $\lambda_S$ is expected to face tighter constraints, with the allowed regions around the Higgs resonance shrinking further. Outside of this Higgs resonance region, a substantial portion of the viable parameter space in the model can be probed by future DM direct detection experiments, including XENONnT/LZ (dashed blue line) and DARWIN (dashed green line).

However, in cases involving new $Z_2$-odd mediators, the real scalar DM-proton scattering cross-sections are suppressed at the loop level. This occurs in processes where $H^\pm$ or $A$ contribute to one-loop box type diagrams involving $W/Z$ bosons, as well as in one-loop Higgs vertex correction diagrams.
As a result, there are parameter spaces that remain below the expected sensitivity of conventional DM direct detection experiments.

We also examine the properties of DM annihilation in indirect detection. 
In the right panel of Fig.~\ref{Fig:exp2}, we display the $95\%$ C.L. regions for the DM annihilation cross-section $\langle \sigma v\rangle$ in the present Universe as a function of DM mass $m_S$.
In the case of DM mass $m_S\approx62.5\,\text{GeV}$, the annihilation process dominated by the Higgs resonance mechanism exhibits the cross-section at the present time  significantly greater than the thermal value of $\langle \sigma v\rangle \sim 10^{-26}\,\rm{cm^3 s^{-1}}$.
On the other hand, we have also confirmed that Higgs resonance and $SS\to WW^*$ annihilation mechanisms can satisfy the correct DM relic density while providing a natural explanation for the GCE and AMS-02 $\overline{p}$ excess.
This corresponds to DM masses around $m_S \approx 62.5\,\text{GeV}$ and $m_S \approx 72\,\text{GeV}$, respectively.

Finally, we consider several LHC search strategies within this model, including mono-jet, compressed mass spectrum, and other SUSY-like searches, and recast these analyses to impose 
constraints on the model parameters. For mono-jet searches, the signal signature consists of an energetic jet and significant missing transverse momentum. Depending on whether $n_1$ is decoupled or not, we specifically classify the following two scenarios: 
\begin{enumerate}
\item Decoupled $n_1$: This scenario is similar to the case in the inert 2HDM, and the signal process is denoted as $pp\to SS+$ jets. We apply the most restrictive cut-window "IM12" in Table.6 of Ref.~\cite{ATLAS:2021kxv} from the ATLAS collaboration, which under yields $207$ signal events and $223\pm 19$ background events and recast this inclusive signal region via \texttt{MadAnalysis5}~\cite{Dumont:2014tja}. 
 We can obtain relevant constraints on the coupling $\lambda_S$ for various DM masses $m_S$. For $45$ GeV $\lesssim m_S\lesssim \frac{m_h}{2}$, the $\lambda_S$ values are excluded in the range from $2.15\times 10^{-2}$ to $4.20\times 10^{-2}$. As $m_S$ increases, the constraints on $\lambda_S$ become less stringent. For $\frac{m_h}{2}\lesssim m_S\lesssim 80$ GeV, the mediator, Higgs boson, can only be produced off-shell, leading to even weaker constraints on $\lambda_S$, ranging from $0.258$ to $3.568$.
\item Non-decoupled $n_1$: In this scenario, in addition to the process above, the process $pp\to n_1\overline{n_1}+$ jets with $n_1\to\nu_l S$ and $\overline{n_1}\to\overline{\nu_l} S$ is also involved. We found that the cross section for this process is relatively small, approximately $10^{-2}$ fb, even assuming a branching ratio of $n_1\to\nu_l S$ and $\overline{n_1}\to\overline{\nu_l} S$ of $100\%$. As a result, $y_{H1}$ in the perturbative region cannot be effectively constrained. Nevertheless, since these two processes cannot be distinguished these in the ATLAS detectors, their contributions are combined in this scenario. To obtain a stronger constraint on $\lambda_S$, we set $\Delta M_{\Psi\chi} = 1$ GeV with $m_{n_1} = 100$ GeV (minimum mass) and $y_{H_1} = \sqrt{4\pi}$ (maximum coupling). For $45$ GeV $\lesssim m_S\lesssim \frac{m_h}{2}$ and $\frac{m_h}{2}\lesssim m_S\lesssim 80$ GeV, the $\lambda_S$ values are excluded in the ranges of $2.16\times 10^{-2}$ to $3.70\times 10^{-2}$ and $0.11$ to $3.24$, respectively. 
\end{enumerate}  
Compared to the left panel of Fig.~\ref{Fig:couplings}, we find that the current mono-jet searches from Ref.~\cite{ATLAS:2021kxv} still cannot exclude the allowed parameter space for either of these two scenarios.         

\begin{figure}
  \centering
  \includegraphics[width=16.4cm,height=8.1cm]{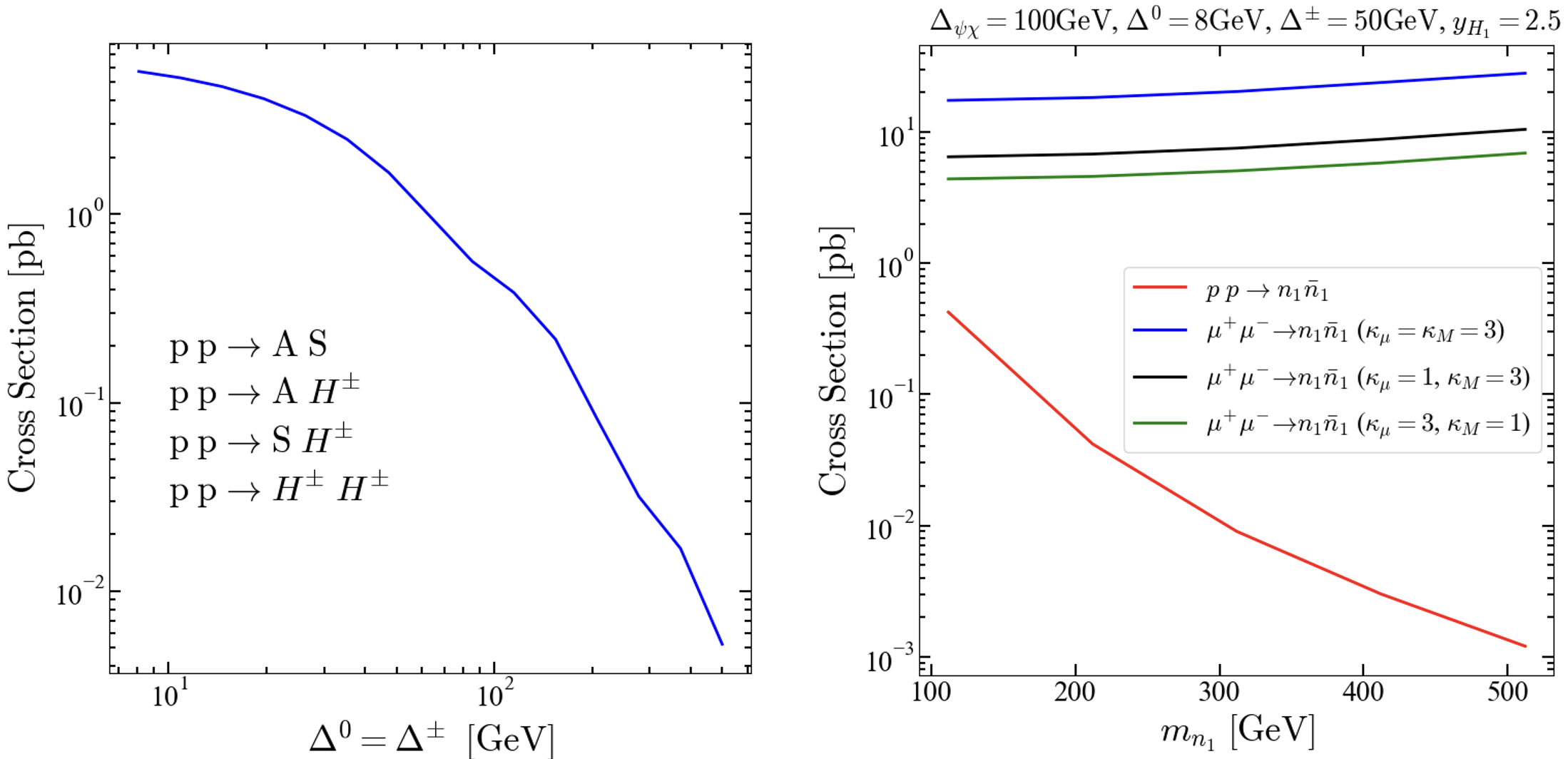}
  \caption{Left panel: The total production cross section, including $pp \to AS$, $pp \to AH^{\pm}$, $pp \to SH^{\pm}$, and $pp \to H^{\pm}H^{\mp}$, is calculated with DM mass $m_S = 60$ GeV while varying the mass splitting $\Delta^{0,\pm}$ at $\sqrt{s}=13$ TeV.  Right panel: The production cross section of $p p \to n_1 \overline{n_1}$ at $\sqrt{s}=13$ TeV and $\mu^{+}\mu^{-} \to n_1 \overline{n_1}$ at $\sqrt{s}=3$ TeV with varying the mass of $n_1$.}
  \label{Fig:cs_n1}
\end{figure}

\begin{figure}
  \centering
 \includegraphics[width=1.0\textwidth]{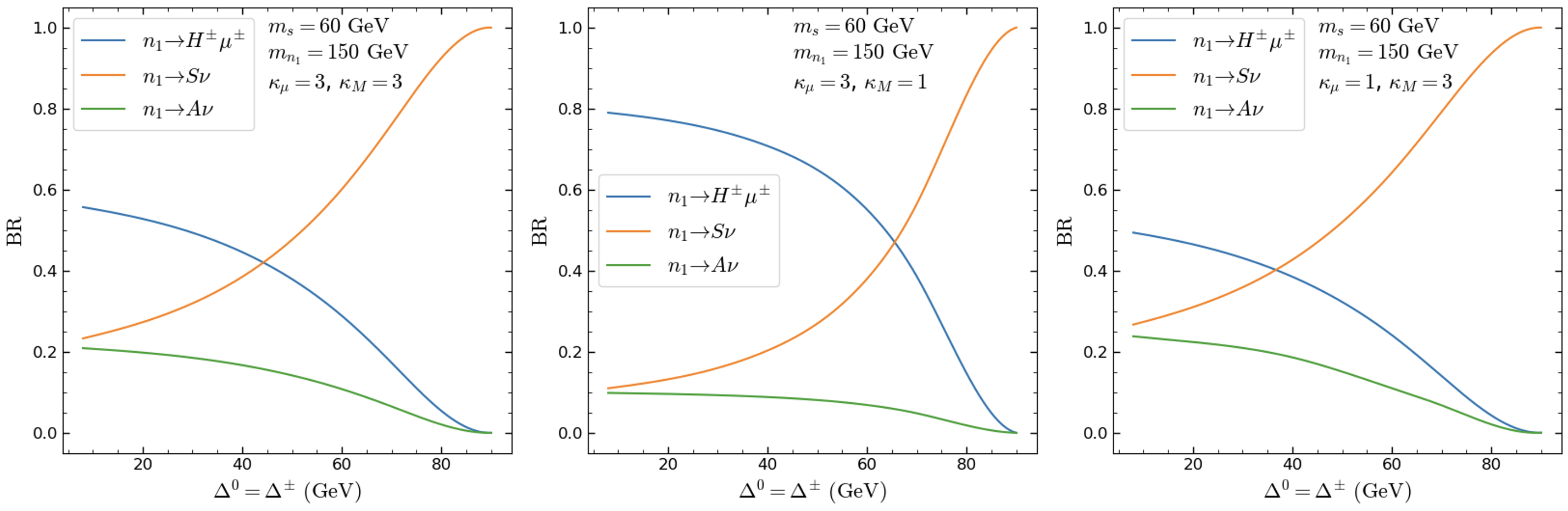}
  \caption{The decay branching ratios of $n_1$ with  $\Delta^0 =\Delta^{\pm}$, $m_s=60$ GeV, and $m_{n_1}=150$ GeV. Left panel: The scenario where $\kappa_{M}=\kappa_{\mu}=3$; Central panel: The case with $\kappa_{\mu}=3$ and $\kappa_{M}=1$; Right panel: The configuration with $\kappa_{\mu}=1$ and $\kappa_{M}=3$.}
  \label{Fig:Brn1}
\end{figure}

We then turn to compressed mass spectrum searches. For small mass splitting $\Delta^{0,\pm}$, the signal signatures typically contain soft objects. For example, searches for SUSY particles in final states with two or three soft leptons and missing transverse momentum have been extensively conducted by the CMS~\cite{CMS:2021edw} and ATLAS~\cite{ATLAS:2019lng} collaborations. As in the mono-jet analysis, we classify the signals into two scenarios depending on whether $n_1$ is decoupled or not: 
\begin{enumerate}
\item Decoupled $n_1$: In this scenario, the heavier neutral scalar $A$ produced at the LHC can decay into a pair of leptons, with the invariant mass of these leptons being sensitive to the mass splitting $\Delta^0$. We recast the current exclusion limit from the CMS SUSY compressed mass spectrum searches~\cite{CMS:2021edw} via MadAnalysis5. Specifically, we focus on the exclusive signal region, $2l$-Ewk SR with $p^{\text{miss}}_T = 200-240$ [GeV] and $M(ll) = 1-4$ [GeV] as outlined in Table 4 of Ref.~\cite{CMS:2021edw}. In this region, $2$  events were observed, while $5.5\pm 2.5$ total background events were expected. This exclusive signal region provides the most stringent constraints. We consider all possible signal processes that produce soft leptons signatures, including $SA$, $AH^{\pm}$, and $H^{\pm}H^{\mp}$. As shown in the left panel of Fig.~\ref{Fig:cs_n1}, the combined cross sections for these processes, without the decays of $A$ and $H^{\pm}$, are on the order of a few pb. Moreover, the decay branching ratios of $H^{\pm}$ are approximately $BR(H^{\pm}\to Sjj')\sim 0.67$ and $BR(H^{\pm}\to Sl^{\pm}\nu)\sim 0.33$. Similarly, the branching ratios for $A$ are $BR(A\to Sjj)\sim 0.66$, $BR(A\to Sl^{+}l^{-})\sim 0.12$, and $BR(A\to S\nu\overline{\nu})\sim 0.22$. In the parameter space of interest, these branching ratios show slight dependence on the mass splitting $\Delta^{0,\pm}$. 
However, these cross sections are still not large enough to exclude the parameter space for the dominant $SA$ co-annihilaiton region, except for $m_S\lesssim 60$ GeV with $\Delta^0 < 10$ GeV, as indicated by the black solid line in the upper-right panel of  Fig.~\ref{Fig:ms_mH}. Note that $\Delta^{\pm} = 50$ GeV is fixed in this analysis\footnote{We observe that the signal efficiency slightly increases for enhanced $\Delta^{\pm}$ when $m_S \gtrsim 70$ GeV. However, these effects are minimal, and no allowed parameter space exists in those regions.}. The magnitude of the mass splitting inversely correlates with the stringency of the imposed constraints. Specifically, larger mass splittings result in less stringent restrictions, while smaller mass splittings enforce stricter constraints. However, the overall impact on the exclusion line remains relatively modest. As we will see in Sec.~\ref{sec:future}, the upcoming luminosity upgrade of the LHC will thoroughly test the entire $SA$ co-annihilation region.     
\item Non-decoupled $n_1$: In this scenario, we first consider the new process of generating soft leptons from $n_1\overline{n_1}$ production. Both $n_1$ and $\overline{n_1}$ subsequently decay either directly to $S$ or undergo a cascade decay via $A$ and/or $H^{\pm}$, yielding soft leptons plus missing transverse momentum. We demonstrate the production cross section for the process $pp\to n_1\overline{n_1}$ in the right panel of Fig.~\ref{Fig:cs_n1} and find that it decreases notably with increasing $m_{n_1}$. Here $y_{H_1}=2.5$, $\Delta_{\Psi\chi} = 100$ GeV, $\kappa_{\mu}=\kappa_{M}=0.1$ are fixed. Moreover, the decay modes of $n_1$ depend on the mass splitting $\Delta^{0,\pm}$, and we find that the decay channel $n_1\to S\nu_l$ becomes dominant for the compressed mass spectrum, as shown in Fig.~\ref{Fig:Brn1}. Here, we fixed $m_s = 60$ GeV and $m_{n_1} = 150$ GeV for the demonstration purpose. Therefore, we conclude that the cross sections for this process are considerably small and lies far beyond the current detection capabilities at the LHC. In the end, although we combine these two types of processes in this scenario for the recasting, the exclusion regions are almost the same as those shown in the upper-right panel of Fig.~\ref{Fig:ms_mH}.
\end{enumerate}

Additionally, we investigate the $SA$, $AH^{\pm}$, and $H^{\pm}H^{\mp}$ production processes with larger mass splitting $\Delta^{0,\pm}$ to generate 2-leptons ($2l$) or 3-leptons ($3l$) final states. Using MadAnalysis5, we recast the analyses from the ATLAS and CMS collaborations~\cite{ATLAS:2019lff,ATLAS:2019wgx,CMS:2020bfa}, specifically focusing on the ranges $100$ GeV $\lesssim\Delta^{0,\pm} < 500$ GeV. Unfortunately, our findings indicate that the larger mass splitting regions of $\Delta^{0,\pm}$ remain largely unexplored due to the very small cross sections shown in the left panel of Fig.~\ref{Fig:cs_n1}. As a result, a significant portion of the parameter space remains untested with the current LHC data.

\subsection{Future predictions at collider experiments}
\label{sec:future}

We further employ the rescaling method to estimate future bounds in this model at the LHC with increased luminosity. The detailed calculation procedure is as follows: We first estimate the future signal events ($s$) and background events ($b$) as 
\beqa
s = \sigma_f \varepsilon \mathcal{L}_f  \;,
\eeqa 

\beqa
b = b_c \times \frac{\mathcal{L}_f}{\mathcal{L}_c}  \;,
\eeqa  
where $\sigma_f$ represents the future signal cross section, and $\epsilon$ denotes the signal efficiency, which we assume remains the same as in the current analysis obtained from the recasting using MadAnalysis5. The future luminosity is denoted by $\mathcal{L}_f$. Here, $b_c$ and $\mathcal{L}_c$ represent the current background events and luminosity, respectively. Next, the statistical significance of the observed signal is calculated using the following formula:
\beqa
Z = \sqrt{2\cdot\left[(s+b)\cdot ln\left(\frac{(s+b)(b+\sigma^2_b)}{b^2 +(s+b)\sigma^2_b}\right)-\frac{b^2}{\sigma^2_b}\cdot ln\left(1+\frac{\sigma^2_b s}{b(b+\sigma^2_b)}\right)\right]} \;,
\label{eq:uncertainty}
\eeqa 
where $\sigma_b$ represents the estimated systematic uncertainty of the background events. A statistical significance corresponding to a $95\%$ confidence level ($Z=1.96$), is used to estimate the future bounds.

In the decoupled $n_1$ scenario, the signal cross sections are proportional to the square of the $\lambda_S$ values. Consequently, we can translate future bonds on $\sigma_f$ to constraints on $\lambda_S$. As examples, for three different assumptions regarding the systematic uncertainty of the background events at the High-Luminosity LHC (${\cal L} = 3000$ fb$^{-1}$), the estimated future bounds for $m_S = 55$ GeV are follows: $\lambda_S\lesssim 6\times 10^{-3}$ for $\sigma_b /b\sim 0$, $\lambda_S\lesssim 0.018$ for $\sigma_b /b = 10\%$, and $\lambda_S\lesssim 0.026$ for $\sigma_b /b = 20\%$. These results indicate that level of systematic uncertainties becomes a critical factor in predicting future bounds on $\lambda_S$. Using the same method as described above, we also analyze the mono-jet signature in the non-decoupled $n_1$ scenario at the LHC with ${\cal L} = 3000$ fb$^{-1}$. However, even with increased luminosity, the constraint on $ |y_{H_1}| $ remains difficult to achieve below $\sqrt{4\pi}$. For instance, with $m_S = 55$ GeV, $m_{n_1} = 100$ GeV, and $\Delta_{\psi\chi} = 1$ GeV, the required luminosity to constrain $ |y_{H_1}| < \sqrt{4\pi}$ is approximately $3440$ fb$^{-1}$ without involving systematic uncertainties.

For the analysis of compressed mass spectra, we also estimate the future bounds at a $95\%$ confidence level for two scenarios under increased luminosity: 
\begin{enumerate}
\item Decoupled $n_1$: 
In this scenario, significant constraints have already been imposed with the current LHC luminosity. We set the future luminosity to $\mathcal{L}$=300$fb^{-1}$ for the Run-3 LHC. Given the small number of observed events and the relatively large uncertainties in background events within the exclusive signal region, $2l$-Ewk SR with $p^{\text{miss}}_T = 200-240$ [GeV] and $M(ll) = 1-4$ [GeV] as outlined in Table 4 of Ref.~\cite{CMS:2021edw}, we conservatively assume a maximum systematic uncertainty of $\sigma_b /b = 45\%$ for the SM background estimation. Under these conditions, we find that the entice parameter space can be excluded, as shown by the green dashed line in the upper-right panel of Fig.~\ref{Fig:ms_mH}. 
 
\item Non-decoupled $n_1$:
In contract, we find that even with a future luminosity of $3000$ fb$^{-1}$, the parameter $|y_{H_1}|$ cannot be effectively constrained to below $\sqrt{4\pi}$ in this scenario. For instance, taking $m_S=55$ GeV, $\Delta_{\psi\chi}=1$ GeV, $\Delta^{0}=\Delta^{\pm} = 8$ GeV, and $m_{n_1}=100$ GeV as an example, the required luminosity to constrain $|y_{H_1}| < \sqrt{4\pi}$ is approximately $\mathcal{L}$= 3850$fb^{-1}$, without taking into account systematic uncertainties. 
\end{enumerate} 

\begin{figure}
    \centering
    \begin{subfigure}{0.35\textwidth}
        \centering
        \includegraphics[width=\textwidth]{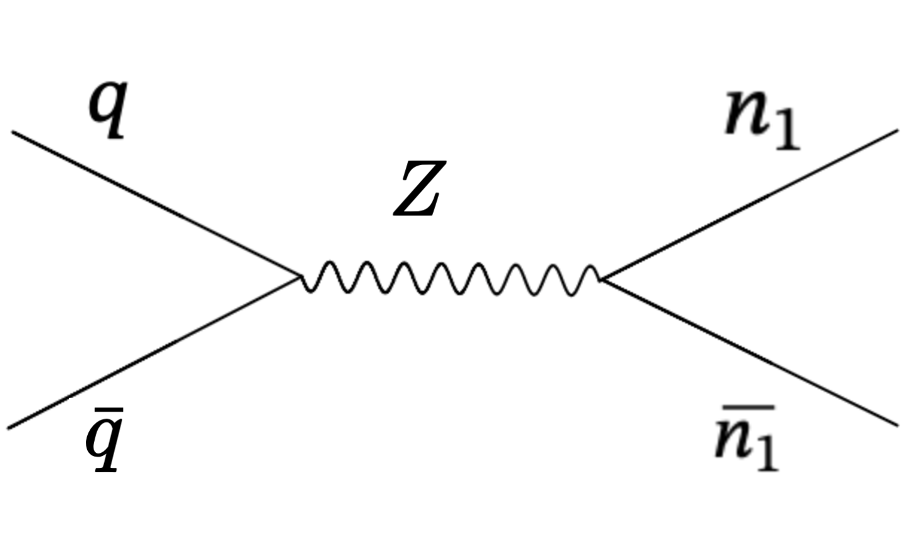}
        \caption{}
        \label{fig:pp3}
    \end{subfigure}
    \hspace{0.02\textwidth} % 调整间距
    \begin{subfigure}{0.35\textwidth}
        \centering
        \includegraphics[width=\textwidth]{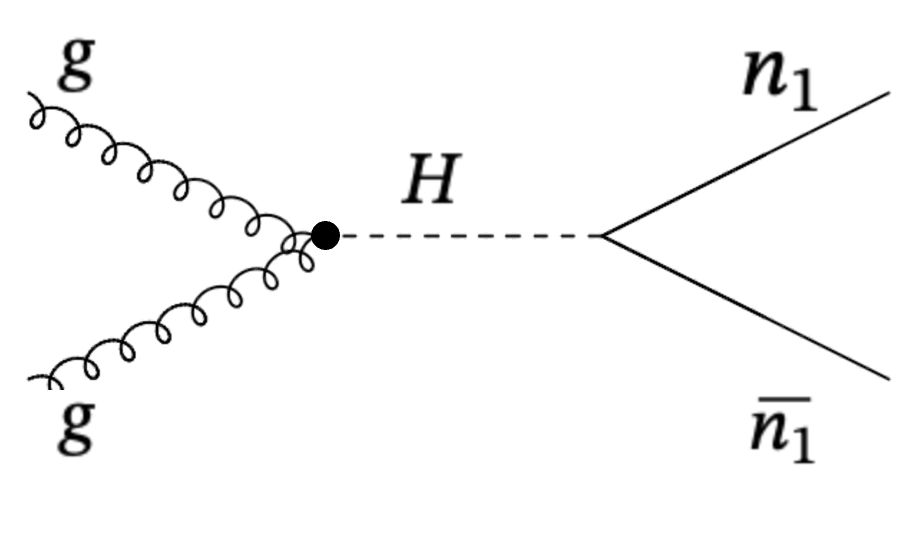}
        \caption{}
        \label{fig:pp4}
    \end{subfigure}
    \hspace{0.02\textwidth} % 调整间距
    \begin{subfigure}{0.35\textwidth}
        \centering
        \includegraphics[width=\textwidth]{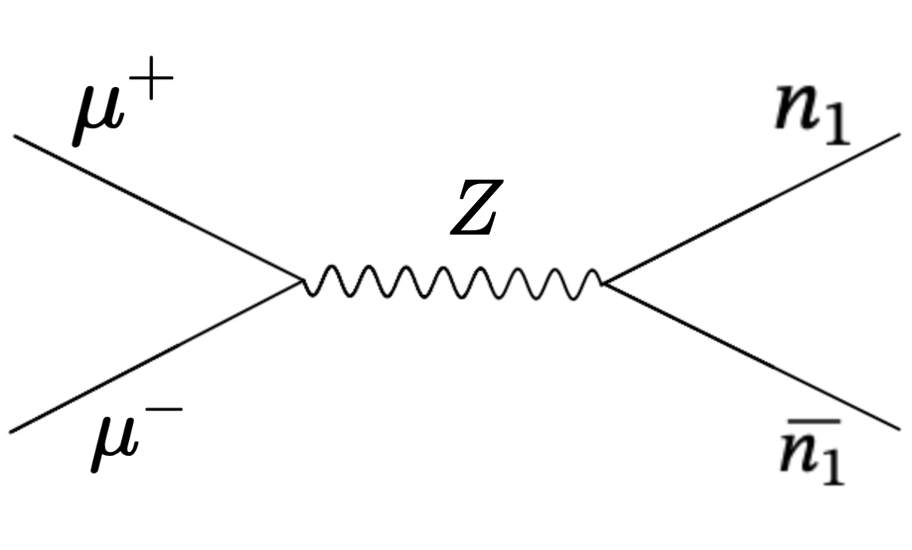}
        \caption{}
        \label{fig:pp5}
    \end{subfigure} 
    \hspace{0.02\textwidth} % 调整间距
    \begin{subfigure}{0.35\textwidth}
        \centering
        \includegraphics[width=\textwidth]{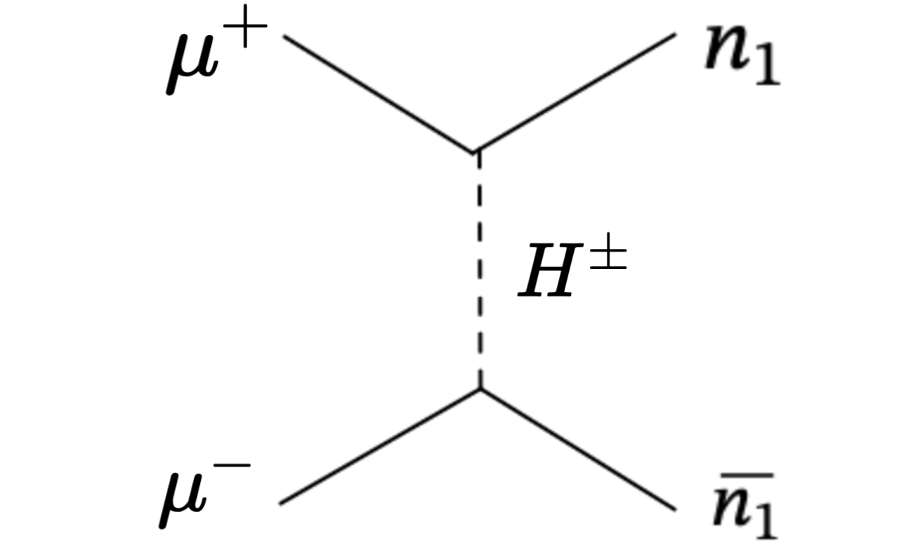}
        \caption{}
        \label{fig:pp5}
    \end{subfigure}
    \caption{Feynman diagrams illustrating the $p p\to n_1 \overline{n_1}$ are shown in subfigures (a) and (b), while subfigures (c) and (d) represent the Feynman diagram for the  $\mu^+\mu^-\to n_1 \overline{n_1}$.}
    \label{fig:fymn}
\end{figure}

The small cross sections of the non-decoupled $n_1$ in compressed mass spectra make them difficult to detect effectively at the LHC, partly due to the decay branching ratios of $n_1$. To address this, we studied the relationship between the $n_1$ decay branching ratios and the mass splittings, $\Delta^0 = \Delta^{\pm}$, under different coupling values of $\kappa_{M}$ and $\kappa_{\mu}$, as shown in Fig.~\ref{Fig:Brn1}. Our findings demonstrate that simultaneous variations in $\kappa_{M}$ and $\kappa_{\mu}$ do not affect the overall decay branching ratio (left panel). Specifically, while $\kappa_{M}$ governs interactions with $\mu^{\pm}_L$ and $\nu^{\pm}$, and $\kappa_{\mu}$ is exclusively related to $\mu^{\pm}_R$, setting $\kappa_{M}=1$ and $\kappa_{\mu}=3$ (central panel) leads to a significant increase in the branching ratio of the $n_1 \to H^{\pm} \mu^{\mp}$ decay channel. Conversely, when $\kappa_{M}=3$ and $\kappa_{\mu}=1$ (right panel), there is a marked increase in the branching ratios of the $n_1 \to A \nu$ and $n_1 \to S \nu$ decay channels. Thus, studying the decay branching ratios of $n_1$ offers insight into the relative magnitudes of the couplings $\kappa_{M}$ and $\kappa_{\mu}$.  

Finally, since the future bounds for this model at the High-Luminsity LHC remain limited, our study further explores the potential of high-energy muon colliders. With a center-of-mass energy of $3$ TeV~\cite{MuonCollider:2022xlm,Black:2022cth}, these next-generation facilities could significantly imporve the prospects of detecting the $n_1 \overline{n_1}$ production process. The key difference lies in the interaction type: the $\mu^+\mu^-\to n_1 \overline{n_1}$ process primarily occurs via $t$-channel $H^{\pm}$ exchange, while the $p p\to n_1 \overline{n_1}$ process is dominated by the Drell-Yan process. The relevant Feynman diagrams for these processes are shown in Fig.~\ref{fig:fymn}. As illustrated in the right panel of Fig.~\ref{Fig:cs_n1}, the cross sections for $\mu^{+}\mu^{-} \to n_1 \overline{n_1}$ with couplings $\kappa_{\mu} = \kappa_M = 3$ (orange line) significantly surpass those of the $p p\to n_1 \overline{n_1}$ process (blue line), and crucially, do not decrease as $m_{n_1}$ increases. In this figure, we fixed $\Delta_{\psi\chi} = 100$ GeV, $\Delta^0 = 8$ GeV, $\Delta^{\pm} = 50$ GeV, and $y_{H_1} = 2.5$ for the demonstration purpose. Additionally, we varied the ratios of couplings $\kappa_{\mu}$ and $\kappa_M$, with $\kappa_{\mu} = 1$ and $\kappa_M = 3$ (green line), and $\kappa_{\mu} = 3$ and $\kappa_M = 1$ (red line), for the $\mu^+\mu^-\to n_1 \overline{n_1}$ process, as shown in the right panel of Fig.~\ref{Fig:cs_n1}. It is clear that the production cross sections in this process are sensitive to the relative magnitudes of $\kappa_{\mu}$ and $\kappa_M$, similar to their influence on $n_1$ decay branching ratios. Thus, a future muon collider could be an ideal platform for detecting signals of this process.

\section{Conclusion}
\label{sec:conclude}

In this work, we present a DM model that extends the particle SM by incorporating singlet and doublet $Z_2$-odd Dirac fermion fields, alongside a $Z_2$-odd doublet complex scalar field. This simple DM model holds the capacity to elucidate the excesses observed in the precision measurements of the muon magnetic moment and the $W$ boson mass from Fermilab experiments, along with addressing the Galactic Center GeV excess (GCE) and the AMS-02 antiproton ($\overline{p}$) excess.

We summarize the characteristics of the feasible parameter regions as follows: First, the mass of the scalar DM within the range of $53$ to $74$ GeV remains allowed, regardless of the anomalies in the muon magnetic moment and the $W$ boson mass, as shown in Fig.~\ref{Fig:ms_mH}. Second, both the inert Higgs sector and the new Dirac fermions contribute to the $W$ boson mass as illustrated in Fig.~\ref{fig:feynman_mw}. Therefore, a smaller charged mass splitting, $\Delta^{\pm}$, is still permissible for explaining the $W$ boson mass excess, which is notable difference from the inert two-Higgs-doublet models~\cite{Fan:2022dck}. Third, the co-annihilation dominant regions with neutral mass splitting $\Delta^0\lesssim 10$ GeV are strongly constrained by current and forthcoming LHC searches, as indicated in Fig.~\ref{Fig:ms_mH}. Fourth, although the recent LZ results exclude a substantial portion of the parameter space around the Higgs resonance, smaller $\lambda_S$ values are still allowed and remain challenging to probe in future DM direct detection experiments. Fifth, there are small regions for DM masses around $63$ GeV and $71-73$ GeV that can simultaneously explain both the GCE and the AMS-02 $\overline{p}$ excess, as seen in the right panel of Fig.~\ref{Fig:exp2}. Finally, the model parameters $y_{H_1}$, $k_M$, and $k_{\mu}$, along with the neutral dark fermion mass, are crucial for explaining the muon magnetic moment excess, as shown in the right panel of Fig.~\ref{Fig:couplings}. However, these parameters are difficult to examine at the LHC. Fortunately, a high-energy muon collider may provide an ideal environment to detect them, potentially confirming or ruling out the predictions in this model.

Before closing, we would like to offer the following remarks for this DM model: (1) While our focus has been on selecting the scalar field $S$ as the DM candidate in this study, it is worth noting that the fermion field $n_1$ could also serve as a DM candidate when $m_{n_1} < m_S$. This intriguing possibility remains open for exploration in future investigations~\cite{Bhattacharya:2018fus}. (2) The multi-muons plus missing energy serves as a common signature arising from processes involving $\psi^{\pm}$ and $n_2$~\cite{Bissmann:2020lge,ATLAS:2021yyr,Lu:2021vcp}. These distinctive signals warrant investigation at high luminosity large hadron colliders and potential future muon colliders. (3) Similar to the inert two Higgs doublet model~\cite{Blinov:2015vma} and the model incorporating vector-like leptons and scalar singlets~\cite{Bell:2019mbn}, which are capable of inducing strong first-order phase transitions (SFOPT), our model also holds the potential to trigger SFOPT. This possibility invites further study and exploration in the future.

%%%%%%%%%%%%%%%%%%%%%%%%%%%%%%%%%%%%%%%%%%%%%
\section*{Acknowledgments}
%%%%%%%%%%%%%%%%%%%%%%%%%%%%%%%%%%%%%%%%%%%%%
This work was supported by the National Key Research and Development Program of China (No. 2022YFF0503304), and the
Project for Young Scientists in Basic Research of the Chinese Academy of Sciences (No. YSBR-092). 
CTL and XYL are supported by the National Natural Science Foundation of China (NNSFC) under grant No.~12335005 and the Special funds for postdoctoral overseas recruitment, Ministry of Education of China (No.~164080H0262403).
TPT is supported by the Jiangsu Province Post Doctoral Foundation (No. 2024ZB713).
This work was also supported in part by the NNSFC under grant Nos. 19Z103010239 and 12350410369, and Vietnam National Foundation for Science and Technology Development (NAFOSTED) under grant number 103.01-2023.50 (VQT). VQT would like to thank the Medium and High Energy Physics group at the Institute of Physics, Academia Sinica for their hospitality during the course of this work.

%%%%%%%%%%%%%%%%%%%%%%%%%%%%%%%%%%%%%%%%%%%%%%%%%%%%


\begin{thebibliography}{99}
%\cite{Muong-2:2023cdq}
\bibitem{Muong-2:2023cdq}
D.~P.~Aguillard \textit{et al.} [Muon g-2],
%``Measurement of the Positive Muon Anomalous Magnetic Moment to 0.20~ppm,''
Phys. Rev. Lett. \textbf{131}, no.16, 161802 (2023)
doi:10.1103/PhysRevLett.131.161802
[arXiv:2308.06230 [hep-ex]].
%345 citations counted in INSPIRE as of 27 Sep 2024

%\cite{Aoyama:2020ynm}
\bibitem{Aoyama:2020ynm}
T.~Aoyama, N.~Asmussen, M.~Benayoun, J.~Bijnens, T.~Blum, M.~Bruno, I.~Caprini, C.~M.~Carloni Calame, M.~C\`e and G.~Colangelo, \textit{et al.}
%``The anomalous magnetic moment of the muon in the Standard Model,''
Phys. Rept. \textbf{887}, 1-166 (2020)
doi:10.1016/j.physrep.2020.07.006
[arXiv:2006.04822 [hep-ph]].
%1428 citations counted in INSPIRE as of 27 Sep 2024

%\cite{Borsanyi:2020mff}
\bibitem{Borsanyi:2020mff}
S.~Borsanyi, Z.~Fodor, J.~N.~Guenther, C.~Hoelbling, S.~D.~Katz, L.~Lellouch, T.~Lippert, K.~Miura, L.~Parato and K.~K.~Szabo, \textit{et al.}
%``Leading hadronic contribution to the muon magnetic moment from lattice QCD,''
Nature \textbf{593}, no.7857, 51-55 (2021)
doi:10.1038/s41586-021-03418-1
[arXiv:2002.12347 [hep-lat]].
%929 citations counted in INSPIRE as of 26 Sep 2024

%\cite{CDF:2022hxs}
\bibitem{CDF:2022hxs}
T.~Aaltonen \textit{et al.} [CDF],
%``High-precision measurement of the $W$          boson mass with the CDF II detector,''
Science \textbf{376}, no.6589, 170-176 (2022)
doi:10.1126/science.abk1781
%633 citations counted in INSPIRE as of 25 Sep 2024

%\cite{ATLAS:2023fsi}
\bibitem{ATLAS:2023fsi}
 [ATLAS],
%``Improved W boson Mass Measurement using 7 TeV Proton-Proton Collisions with the ATLAS Detector,''
ATLAS-CONF-2023-004.
%51 citations counted in INSPIRE as of 26 Sep 2024

%\cite{CMS:2024}
\bibitem{CMS:2024}
 [CMS],
%``Measurement of the W boson mass in proton-proton collisions at sqrts = 13 TeV,''
CMS-PAS-SMP-23-002.

%\cite{Roos:2010wb}
\bibitem{Roos:2010wb}
M.~Roos,
%``Dark Matter: The evidence from astronomy, astrophysics and cosmology,''
[arXiv:1001.0316 [astro-ph.CO]].
%92 citations counted in INSPIRE as of 25 Sep 2024

%\cite{Arcadi:2017kky}
\bibitem{Arcadi:2017kky}
G.~Arcadi, M.~Dutra, P.~Ghosh, M.~Lindner, Y.~Mambrini, M.~Pierre, S.~Profumo and F.~S.~Queiroz,
%``The waning of the WIMP? A review of models, searches, and constraints,''
Eur. Phys. J. C \textbf{78}, no.3, 203 (2018)
doi:10.1140/epjc/s10052-018-5662-y
[arXiv:1703.07364 [hep-ph]].
%848 citations counted in INSPIRE as of 26 Sep 2024

%\cite{Bauer:2017qwy}
\bibitem{Bauer:2017qwy}
M.~Bauer and T.~Plehn,
%``Yet Another Introduction to Dark Matter: The Particle Physics Approach,''
Lect. Notes Phys. \textbf{959}, pp. (2019)
Springer, 2019,
doi:10.1007/978-3-030-16234-4
[arXiv:1705.01987 [hep-ph]].
%112 citations counted in INSPIRE as of 25 Sep 2024

%\cite{Schumann:2019eaa}
\bibitem{Schumann:2019eaa}
M.~Schumann,
%``Direct Detection of WIMP Dark Matter: Concepts and Status,''
J. Phys. G \textbf{46}, no.10, 103003 (2019)
doi:10.1088/1361-6471/ab2ea5
[arXiv:1903.03026 [astro-ph.CO]].
%510 citations counted in INSPIRE as of 25 Sep 2024

%\cite{Slatyer:2021qgc}
\bibitem{Slatyer:2021qgc}
T.~R.~Slatyer,
%``Les Houches Lectures on Indirect Detection of Dark Matter,''
SciPost Phys. Lect. Notes \textbf{53}, 1 (2022)
doi:10.21468/SciPostPhysLectNotes.53
[arXiv:2109.02696 [hep-ph]].
%49 citations counted in INSPIRE as of 25 Sep 2024

%\cite{Boveia:2018yeb}
\bibitem{Boveia:2018yeb}
A.~Boveia and C.~Doglioni,
%``Dark Matter Searches at Colliders,''
Ann. Rev. Nucl. Part. Sci. \textbf{68}, 429-459 (2018)
doi:10.1146/annurev-nucl-101917-021008
[arXiv:1810.12238 [hep-ex]].
%221 citations counted in INSPIRE as of 25 Sep 2024

%\cite{LUX-ZEPLIN:2022xrq}
\bibitem{LUX-ZEPLIN:2022xrq}
J.~Aalbers \textit{et al.} [LZ],
%``First Dark Matter Search Results from the LUX-ZEPLIN (LZ) Experiment,''
Phys. Rev. Lett. \textbf{131}, no.4, 041002 (2023)
doi:10.1103/PhysRevLett.131.041002
[arXiv:2207.03764 [hep-ex]].
%694 citations counted in INSPIRE as of 27 Sep 2024

%\cite{XENONCollaboration:2023orw}
\bibitem{XENONCollaboration:2023orw}
E.~Aprile \textit{et al.} [XENON],
%``First Dark Matter Search with Nuclear Recoils from the XENONnT Experiment,''
Phys. Rev. Lett. \textbf{131}, no.4, 041003 (2023)
doi:10.1103/PhysRevLett.131.041003
[arXiv:2303.14729 [hep-ex]].
%291 citations counted in INSPIRE as of 27 Sep 2024

%\cite{PerezAdan:2023rsl}
\bibitem{PerezAdan:2023rsl}
D.~Perez Adan [ATLAS and CMS],
%``Dark Matter searches at CMS and ATLAS,''
[arXiv:2301.10141 [hep-ex]].
%8 citations counted in INSPIRE as of 25 Sep 2024

%\cite{Hooper:2010mq}
\bibitem{Hooper:2010mq}
D.~Hooper and L.~Goodenough,
%``Dark Matter Annihilation in The Galactic Center As Seen by the Fermi Gamma Ray Space Telescope,''
Phys. Lett. B \textbf{697}, 412-428 (2011)
doi:10.1016/j.physletb.2011.02.029
[arXiv:1010.2752 [hep-ph]].
%914 citations counted in INSPIRE as of 25 Sep 2024

%\cite{Zhou:2014lva}
\bibitem{Zhou:2014lva}
B.~Zhou, Y.~F.~Liang, X.~Huang, X.~Li, Y.~Z.~Fan, L.~Feng and J.~Chang,
%``GeV excess in the Milky Way: The role of diffuse galactic gamma-ray emission templates,''
Phys. Rev. D \textbf{91}, no.12, 123010 (2015)
doi:10.1103/PhysRevD.91.123010
[arXiv:1406.6948 [astro-ph.HE]].
%189 citations counted in INSPIRE as of 25 Sep 2024

%\cite{Calore:2014xka}
\bibitem{Calore:2014xka}
F.~Calore, I.~Cholis and C.~Weniger,
%``Background Model Systematics for the Fermi GeV Excess,''
JCAP \textbf{03}, 038 (2015)
doi:10.1088/1475-7516/2015/03/038
[arXiv:1409.0042 [astro-ph.CO]].
%543 citations counted in INSPIRE as of 25 Sep 2024

%\cite{Daylan:2014rsa}
\bibitem{Daylan:2014rsa}
T.~Daylan, D.~P.~Finkbeiner, D.~Hooper, T.~Linden, S.~K.~N.~Portillo, N.~L.~Rodd and T.~R.~Slatyer,
%``The characterization of the gamma-ray signal from the central Milky Way: A case for annihilating dark matter,''
Phys. Dark Univ. \textbf{12}, 1-23 (2016)
doi:10.1016/j.dark.2015.12.005
[arXiv:1402.6703 [astro-ph.HE]].
%864 citations counted in INSPIRE as of 25 Sep 2024

%\cite{Cui:2016ppb}
\bibitem{Cui:2016ppb}
M.~Y.~Cui, Q.~Yuan, Y.~L.~S.~Tsai and Y.~Z.~Fan,
%``Possible dark matter annihilation signal in the AMS-02 antiproton data,''
Phys. Rev. Lett. \textbf{118}, no.19, 191101 (2017)
doi:10.1103/PhysRevLett.118.191101
[arXiv:1610.03840 [astro-ph.HE]].
%234 citations counted in INSPIRE as of 25 Sep 2024

%\cite{Cuoco:2016eej}
\bibitem{Cuoco:2016eej}
A.~Cuoco, M.~Kr\"amer and M.~Korsmeier,
%``Novel Dark Matter Constraints from Antiprotons in Light of AMS-02,''
Phys. Rev. Lett. \textbf{118}, no.19, 191102 (2017)
doi:10.1103/PhysRevLett.118.191102
[arXiv:1610.03071 [astro-ph.HE]].
%262 citations counted in INSPIRE as of 25 Sep 2024

%\cite{Cui:2018klo}
\bibitem{Cui:2018klo}
M.~Y.~Cui, X.~Pan, Q.~Yuan, Y.~Z.~Fan and H.~S.~Zong,
%``Revisit of cosmic ray antiprotons from dark matter annihilation with updated constraints on the background model from AMS-02 and collider data,''
JCAP \textbf{06}, 024 (2018)
doi:10.1088/1475-7516/2018/06/024
[arXiv:1803.02163 [astro-ph.HE]].
%41 citations counted in INSPIRE as of 25 Sep 2024

%\cite{Cholis:2019ejx}
\bibitem{Cholis:2019ejx}
I.~Cholis, T.~Linden and D.~Hooper,
%``A Robust Excess in the Cosmic-Ray Antiproton Spectrum: Implications for Annihilating Dark Matter,''
Phys. Rev. D \textbf{99}, no.10, 103026 (2019)
doi:10.1103/PhysRevD.99.103026
[arXiv:1903.02549 [astro-ph.HE]].
%161 citations counted in INSPIRE as of 27 Sep 2024

%\cite{Okada:2014qsa}
\bibitem{Okada:2014qsa}
H.~Okada, T.~Toma and K.~Yagyu,
%``Inert Extension of the Zee-Babu Model,''
Phys. Rev. D \textbf{90}, 095005 (2014)
doi:10.1103/PhysRevD.90.095005
[arXiv:1408.0961 [hep-ph]].
%58 citations counted in INSPIRE as of 25 Sep 2024

%\cite{Peskin:1991sw}
\bibitem{Peskin:1991sw}
M.~E.~Peskin and T.~Takeuchi,
%``Estimation of oblique electroweak corrections,''
Phys. Rev. D \textbf{46}, 381-409 (1992)
doi:10.1103/PhysRevD.46.381
%2744 citations counted in INSPIRE as of 27 Sep 2024

%\cite{Fan:2022dck}
\bibitem{Fan:2022dck}
Y.~Z.~Fan, T.~P.~Tang, Y.~L.~S.~Tsai and L.~Wu,
%``Inert Higgs Dark Matter for CDF II W-Boson Mass and Detection Prospects,''
Phys. Rev. Lett. \textbf{129}, no.9, 091802 (2022)
doi:10.1103/PhysRevLett.129.091802
[arXiv:2204.03693 [hep-ph]].
%114 citations counted in INSPIRE as of 27 Sep 2024

%\cite{Ma:2006km}
\bibitem{Ma:2006km}
E.~Ma,
%``Verifiable radiative seesaw mechanism of neutrino mass and dark matter,''
Phys. Rev. D \textbf{73}, 077301 (2006)
doi:10.1103/PhysRevD.73.077301
[arXiv:hep-ph/0601225 [hep-ph]].
%1527 citations counted in INSPIRE as of 27 Sep 2024

%\cite{Cai:2017jrq}
\bibitem{Cai:2017jrq}
Y.~Cai, J.~Herrero-Garc\'\i{}a, M.~A.~Schmidt, A.~Vicente and R.~R.~Volkas,
%``From the trees to the forest: a review of radiative neutrino mass models,''
Front. in Phys. \textbf{5}, 63 (2017)
doi:10.3389/fphy.2017.00063
[arXiv:1706.08524 [hep-ph]].
%345 citations counted in INSPIRE as of 09 Dec 2024

%\cite{Sokolowska:2011sb}
\bibitem{Sokolowska:2011sb}
D.~Sokolowska,
%``Dark Matter Data and Constraints on Quartic Couplings in IDM,''
[arXiv:1107.1991 [hep-ph]].
%28 citations counted in INSPIRE as of 25 Sep 2024

%\cite{Abe:2015rja}
\bibitem{Abe:2015rja}
T.~Abe and R.~Sato,
%``Quantum corrections to the spin-independent cross section of the inert doublet dark matter,''
JHEP \textbf{03}, 109 (2015)
doi:10.1007/JHEP03(2015)109
[arXiv:1501.04161 [hep-ph]].
%44 citations counted in INSPIRE as of 25 Sep 2024

%\cite{Banerjee:2019luv}
\bibitem{Banerjee:2019luv}
S.~Banerjee, F.~Boudjema, N.~Chakrabarty, G.~Chalons and H.~Sun,
%``Relic density of dark matter in the inert doublet model beyond leading order: The heavy mass case,''
Phys. Rev. D \textbf{100}, no.9, 095024 (2019)
doi:10.1103/PhysRevD.100.095024
[arXiv:1906.11269 [hep-ph]].
%32 citations counted in INSPIRE as of 25 Sep 2024

%\cite{Tsai:2019eqi}
\bibitem{Tsai:2019eqi}
Y.~L.~S.~Tsai, V.~Q.~Tran and C.~T.~Lu,
%``Confronting dark matter co-annihilation of Inert two Higgs Doublet Model with a compressed mass spectrum,''
JHEP \textbf{06}, 033 (2020)
doi:10.1007/JHEP06(2020)033
[arXiv:1912.08875 [hep-ph]].
%23 citations counted in INSPIRE as of 25 Sep 2024

%\cite{ATLAS:2019lff}
\bibitem{ATLAS:2019lff}
G.~Aad \textit{et al.} [ATLAS],
%``Search for electroweak production of charginos and sleptons decaying into final states with two leptons and missing transverse momentum in $\sqrt{s}=13$ TeV $pp$ collisions using the ATLAS detector,''
Eur. Phys. J. C \textbf{80}, no.2, 123 (2020)
doi:10.1140/epjc/s10052-019-7594-6
[arXiv:1908.08215 [hep-ex]].
%346 citations counted in INSPIRE as of 27 Sep 2024

%\cite{CMS:2020bfa}
\bibitem{CMS:2020bfa}
A.~M.~Sirunyan \textit{et al.} [CMS],
%``Search for supersymmetry in final states with two oppositely charged same-flavor leptons and missing transverse momentum in proton-proton collisions at $\sqrt{s} =$ 13 TeV,''
JHEP \textbf{04}, 123 (2021)
doi:10.1007/JHEP04(2021)123
[arXiv:2012.08600 [hep-ex]].
%140 citations counted in INSPIRE as of 25 Sep 2024

%\cite{Calibbi:2018rzv}
\bibitem{Calibbi:2018rzv}
L.~Calibbi, R.~Ziegler and J.~Zupan,
%``Minimal models for dark matter and the muon g\ensuremath{-}2 anomaly,''
JHEP \textbf{07}, 046 (2018)
doi:10.1007/JHEP07(2018)046
[arXiv:1804.00009 [hep-ph]].
%92 citations counted in INSPIRE as of 25 Sep 2024

%\cite{Arcadi:2021cwg}
\bibitem{Arcadi:2021cwg}
G.~Arcadi, L.~Calibbi, M.~Fedele and F.~Mescia,
%``Muon $g-2$ and $B$-anomalies from Dark Matter,''
Phys. Rev. Lett. \textbf{127}, no.6, 061802 (2021)
doi:10.1103/PhysRevLett.127.061802
[arXiv:2104.03228 [hep-ph]].
%77 citations counted in INSPIRE as of 25 Sep 2024

%\cite{Arnan:2019uhr}
\bibitem{Arnan:2019uhr}
P.~Arnan, A.~Crivellin, M.~Fedele and F.~Mescia,
%``Generic Loop Effects of New Scalars and Fermions in $b\to s\ell^+\ell^-$, $(g-2)_\mu$ and a Vector-like $4^{\rm th}$ Generation,''
JHEP \textbf{06}, 118 (2019)
doi:10.1007/JHEP06(2019)118
[arXiv:1904.05890 [hep-ph]].
%95 citations counted in INSPIRE as of 25 Sep 2024

%\cite{Barbieri:2006dq}
\bibitem{Barbieri:2006dq}
R.~Barbieri, L.~J.~Hall and V.~S.~Rychkov,
%``Improved naturalness with a heavy Higgs: An Alternative road to LHC physics,''
Phys. Rev. D \textbf{74}, 015007 (2006)
doi:10.1103/PhysRevD.74.015007
[arXiv:hep-ph/0603188 [hep-ph]].
%1009 citations counted in INSPIRE as of 26 Sep 2024

%\cite{Arhrib:2012ia}
\bibitem{Arhrib:2012ia}
A.~Arhrib, R.~Benbrik and N.~Gaur,
%``$H\to \gamma \gamma$ in Inert Higgs Doublet Model,''
Phys. Rev. D \textbf{85}, 095021 (2012)
doi:10.1103/PhysRevD.85.095021
[arXiv:1201.2644 [hep-ph]].
%225 citations counted in INSPIRE as of 26 Sep 2024

%\cite{Swiezewska:2012ej}
\bibitem{Swiezewska:2012ej}
B.~\'Swie\.zewska,
%``Yukawa independent constraints for two-Higgs-doublet models with a 125 GeV Higgs boson,''
Phys. Rev. D \textbf{88}, no.5, 055027 (2013)
[erratum: Phys. Rev. D \textbf{88}, no.11, 119903 (2013)]
doi:10.1103/PhysRevD.88.055027
[arXiv:1209.5725 [hep-ph]].
%58 citations counted in INSPIRE as of 25 Sep 2024

%\cite{Eriksson:2009ws}
\bibitem{Eriksson:2009ws}
D.~Eriksson, J.~Rathsman and O.~Stal,
%``2HDMC: Two-Higgs-Doublet Model Calculator Physics and Manual,''
Comput. Phys. Commun. \textbf{181}, 189-205 (2010)
doi:10.1016/j.cpc.2009.09.011
[arXiv:0902.0851 [hep-ph]].
%518 citations counted in INSPIRE as of 25 Sep 2024

%\cite{DEramo:2007anh}
\bibitem{DEramo:2007anh}
F.~D'Eramo,
%``Dark matter and Higgs boson physics,''
Phys. Rev. D \textbf{76}, 083522 (2007)
doi:10.1103/PhysRevD.76.083522
[arXiv:0705.4493 [hep-ph]].
%121 citations counted in INSPIRE as of 25 Sep 2024

%\cite{Bhattacharya:2018fus}
\bibitem{Bhattacharya:2018fus}
S.~Bhattacharya, P.~Ghosh, N.~Sahoo and N.~Sahu,
%``Mini Review on Vector-Like Leptonic Dark Matter, Neutrino Mass, and Collider Signatures,''
Front. in Phys. \textbf{7}, 80 (2019)
doi:10.3389/fphy.2019.00080
[arXiv:1812.06505 [hep-ph]].
%39 citations counted in INSPIRE as of 25 Sep 2024

%\cite{ParticleDataGroup:2014cgo}
\bibitem{ParticleDataGroup:2014cgo}
K.~A.~Olive \textit{et al.} [Particle Data Group],
%``Review of Particle Physics,''
Chin. Phys. C \textbf{38}, 090001 (2014)
doi:10.1088/1674-1137/38/9/090001
%9269 citations counted in INSPIRE as of 27 Sep 2024

%\cite{Lundstrom:2008ai}
\bibitem{Lundstrom:2008ai}
E.~Lundstrom, M.~Gustafsson and J.~Edsjo,
%``The Inert Doublet Model and LEP II Limits,''
Phys. Rev. D \textbf{79}, 035013 (2009)
doi:10.1103/PhysRevD.79.035013
[arXiv:0810.3924 [hep-ph]].
%282 citations counted in INSPIRE as of 26 Sep 2024

%\cite{Blinov:2015qva}
\bibitem{Blinov:2015qva}
N.~Blinov, J.~Kozaczuk, D.~E.~Morrissey and A.~de la Puente,
%``Compressing the Inert Doublet Model,''
Phys. Rev. D \textbf{93}, no.3, 035020 (2016)
doi:10.1103/PhysRevD.93.035020
[arXiv:1510.08069 [hep-ph]].
%49 citations counted in INSPIRE as of 25 Sep 2024

%\cite{OPAL:2003nhx}
\bibitem{OPAL:2003nhx}
G.~Abbiendi \textit{et al.} [OPAL],
%``Search for anomalous production of dilepton events with missing transverse momentum in e+ e- collisions at s**(1/2) = 183-Gev to 209-GeV,''
Eur. Phys. J. C \textbf{32}, 453-473 (2004)
doi:10.1140/epjc/s2003-01466-y
[arXiv:hep-ex/0309014 [hep-ex]].
%145 citations counted in INSPIRE as of 25 Sep 2024

%\cite{ATLAS:2019lng}
\bibitem{ATLAS:2019lng}
G.~Aad \textit{et al.} [ATLAS],
%``Searches for electroweak production of supersymmetric particles with compressed mass spectra in $\sqrt{s}=$ 13 TeV $pp$ collisions with the ATLAS detector,''
Phys. Rev. D \textbf{101}, no.5, 052005 (2020)
doi:10.1103/PhysRevD.101.052005
[arXiv:1911.12606 [hep-ex]].
%284 citations counted in INSPIRE as of 27 Sep 2024

%\cite{ATLAS:2023tkt}
\bibitem{ATLAS:2023tkt}
G.~Aad \textit{et al.} [ATLAS],
%``Combination of searches for invisible decays of the Higgs boson using 139 fb\ensuremath{-}1 of proton-proton collision data at s=13 TeV collected with the ATLAS experiment,''
Phys. Lett. B \textbf{842}, 137963 (2023)
doi:10.1016/j.physletb.2023.137963
[arXiv:2301.10731 [hep-ex]].
%85 citations counted in INSPIRE as of 27 Sep 2024

%\cite{ATLAS:2020qdt}
\bibitem{ATLAS:2020qdt}
 [ATLAS],
%``A combination of measurements of Higgs boson production and decay using up to $139$ fb$^{-1}$ of proton--proton collision data at $\sqrt{s}=$ 13 TeV collected with the ATLAS experiment,''
ATLAS-CONF-2020-027.
%124 citations counted in INSPIRE as of 25 Sep 2024

%\cite{ATLAS:2022tnm}
\bibitem{ATLAS:2022tnm}
G.~Aad \textit{et al.} [ATLAS],
%``Measurement of the properties of Higgs boson production at $\sqrt{s} = 13$ TeV in the $H\to\gamma\gamma$ channel using $139$ fb$^{-1}$ of $pp$ collision data with the ATLAS experiment,''
JHEP \textbf{07}, 088 (2023)
doi:10.1007/JHEP07(2023)088
[arXiv:2207.00348 [hep-ex]].
%109 citations counted in INSPIRE as of 27 Sep 2024

%\cite{CMS:2021kom}
\bibitem{CMS:2021kom}
A.~M.~Sirunyan \textit{et al.} [CMS],
%``Measurements of Higgs boson production cross sections and couplings in the diphoton decay channel at $ \sqrt{\mathrm{s}} $ = 13 TeV,''
JHEP \textbf{07}, 027 (2021)
doi:10.1007/JHEP07(2021)027
[arXiv:2103.06956 [hep-ex]].
%161 citations counted in INSPIRE as of 25 Sep 2024

%\cite{Swiezewska:2012eh}
\bibitem{Swiezewska:2012eh}
B.~Swiezewska and M.~Krawczyk,
%``Diphoton rate in the inert doublet model with a 125 GeV Higgs boson,''
Phys. Rev. D \textbf{88}, no.3, 035019 (2013)
doi:10.1103/PhysRevD.88.035019
[arXiv:1212.4100 [hep-ph]].
%147 citations counted in INSPIRE as of 25 Sep 2024

%\cite{ATLAS:2021kxv}
\bibitem{ATLAS:2021kxv}
G.~Aad \textit{et al.} [ATLAS],
%``Search for new phenomena in events with an energetic jet and missing transverse momentum in $pp$ collisions at $\sqrt {s}$ =13  TeV with the ATLAS detector,''
Phys. Rev. D \textbf{103}, no.11, 112006 (2021)
doi:10.1103/PhysRevD.103.112006
[arXiv:2102.10874 [hep-ex]].
%230 citations counted in INSPIRE as of 25 Sep 2024

%\cite{CMS:2021edw}
\bibitem{CMS:2021edw}
A.~Tumasyan \textit{et al.} [CMS],
%``Search for supersymmetry in final states with two or three soft leptons and missing transverse momentum in proton-proton collisions at $ \sqrt{s} $ = 13 TeV,''
JHEP \textbf{04}, 091 (2022)
doi:10.1007/JHEP04(2022)091
[arXiv:2111.06296 [hep-ex]].
%76 citations counted in INSPIRE as of 26 Sep 2024

%\cite{Alwall:2014hca}
\bibitem{Alwall:2014hca}
J.~Alwall, R.~Frederix, S.~Frixione, V.~Hirschi, F.~Maltoni, O.~Mattelaer, H.~S.~Shao, T.~Stelzer, P.~Torrielli and M.~Zaro,
%``The automated computation of tree-level and next-to-leading order differential cross sections, and their matching to parton shower simulations,''
JHEP \textbf{07}, 079 (2014)
doi:10.1007/JHEP07(2014)079
[arXiv:1405.0301 [hep-ph]].
%8808 citations counted in INSPIRE as of 27 Sep 2024

%\cite{Sjostrand:2014zea}
\bibitem{Sjostrand:2014zea}
T.~Sj\"ostrand, S.~Ask, J.~R.~Christiansen, R.~Corke, N.~Desai, P.~Ilten, S.~Mrenna, S.~Prestel, C.~O.~Rasmussen and P.~Z.~Skands,
%``An introduction to PYTHIA 8.2,''
Comput. Phys. Commun. \textbf{191}, 159-177 (2015)
doi:10.1016/j.cpc.2015.01.024
[arXiv:1410.3012 [hep-ph]].
%6468 citations counted in INSPIRE as of 27 Sep 2024

%\cite{deFavereau:2013fsa}
\bibitem{deFavereau:2013fsa}
J.~de Favereau \textit{et al.} [DELPHES 3],
%``DELPHES 3, A modular framework for fast simulation of a generic collider experiment,''
JHEP \textbf{02}, 057 (2014)
doi:10.1007/JHEP02(2014)057
[arXiv:1307.6346 [hep-ex]].
%2990 citations counted in INSPIRE as of 27 Sep 2024

%\cite{Dumont:2014tja}
\bibitem{Dumont:2014tja}
B.~Dumont, B.~Fuks, S.~Kraml, S.~Bein, G.~Chalons, E.~Conte, S.~Kulkarni, D.~Sengupta and C.~Wymant,
%``Toward a public analysis database for LHC new physics searches using MADANALYSIS 5,''
Eur. Phys. J. C \textbf{75}, no.2, 56 (2015)
doi:10.1140/epjc/s10052-014-3242-3
[arXiv:1407.3278 [hep-ph]].
%199 citations counted in INSPIRE as of 25 Sep 2024

%\cite{Belanger:2018ccd}
\bibitem{Belanger:2018ccd}
G.~B\'elanger, F.~Boudjema, A.~Goudelis, A.~Pukhov and B.~Zaldivar,
%``micrOMEGAs5.0 : Freeze-in,''
Comput. Phys. Commun. \textbf{231}, 173-186 (2018)
doi:10.1016/j.cpc.2018.04.027
[arXiv:1801.03509 [hep-ph]].
%479 citations counted in INSPIRE as of 25 Sep 2024

%\cite{Planck:2015ica}
\bibitem{Planck:2015ica}
P.~A.~R.~Ade \textit{et al.} [Planck],
%``Planck 2015 results. XXV. Diffuse low-frequency Galactic foregrounds,''
Astron. Astrophys. \textbf{594}, A25 (2016)
doi:10.1051/0004-6361/201526803
[arXiv:1506.06660 [astro-ph.GA]].
%172 citations counted in INSPIRE as of 25 Sep 2024

%\cite{Agashe:2014kda}
\bibitem{Agashe:2014kda}
K.~A.~Olive \textit{et al.} [Particle Data Group],
%``Review of Particle Physics,''
Chin. Phys. C \textbf{38}, 090001 (2014)
doi:10.1088/1674-1137/38/9/090001
%9269 citations counted in INSPIRE as of 27 Sep 2024

%\cite{Abbiendi:2003ji}
\bibitem{Abbiendi:2003ji}
G.~Abbiendi \textit{et al.} [OPAL],
%``Search for anomalous production of dilepton events with missing transverse momentum in e+ e- collisions at s**(1/2) = 183-Gev to 209-GeV,''
Eur. Phys. J. C \textbf{32}, 453-473 (2004)
doi:10.1140/epjc/s2003-01466-y
[arXiv:hep-ex/0309014 [hep-ex]].
%145 citations counted in INSPIRE as of 25 Sep 2024

%\cite{Aghanim:2018eyx}
\bibitem{Aghanim:2018eyx}
N.~Aghanim \textit{et al.} [Planck],
%``Planck 2018 results. VI. Cosmological parameters,''
Astron. Astrophys. \textbf{641}, A6 (2020)
[erratum: Astron. Astrophys. \textbf{652}, C4 (2021)]
doi:10.1051/0004-6361/201833910
[arXiv:1807.06209 [astro-ph.CO]].
%14983 citations counted in INSPIRE as of 27 Sep 2024

%\cite{ATLAS:2018doi}
\bibitem{ATLAS:2018doi}
 [ATLAS],
%``Combined measurements of Higgs boson production and decay using up to 80 fb$^{-1}$ of proton--proton collision data at $\sqrt{s}=$ 13 TeV collected with the ATLAS experiment,''
ATLAS-CONF-2018-031.
%130 citations counted in INSPIRE as of 25 Sep 2024

%\cite{ParticleDataGroup:2022pth}
\bibitem{ParticleDataGroup:2022pth}
R.~L.~Workman \textit{et al.} [Particle Data Group],
%``Review of Particle Physics,''
PTEP \textbf{2022}, 083C01 (2022)
doi:10.1093/ptep/ptac097
%3962 citations counted in INSPIRE as of 27 Sep 2024

%\cite{Rolke:2004mj}
\bibitem{Rolke:2004mj}
W.~A.~Rolke, A.~M.~Lopez and J.~Conrad,
%``Limits and confidence intervals in the presence of nuisance parameters,''
Nucl. Instrum. Meth. A \textbf{551}, 493-503 (2005)
doi:10.1016/j.nima.2005.05.068
[arXiv:physics/0403059 [physics]].
%715 citations counted in INSPIRE as of 25 Sep 2024

%\cite{Fermi-LAT:2016uux}
\bibitem{Fermi-LAT:2016uux}
A.~Albert \textit{et al.} [Fermi-LAT and DES],
%``Searching for Dark Matter Annihilation in Recently Discovered Milky Way Satellites with Fermi-LAT,''
Astrophys. J. \textbf{834}, no.2, 110 (2017)
doi:10.3847/1538-4357/834/2/110
[arXiv:1611.03184 [astro-ph.HE]].
%656 citations counted in INSPIRE as of 25 Sep 2024

%\cite{Cuoco:2016eej}
\bibitem{Cuoco:2016eej}
A.~Cuoco, M.~Kr\"amer and M.~Korsmeier,
%``Novel Dark Matter Constraints from Antiprotons in Light of AMS-02,''
Phys. Rev. Lett. \textbf{118}, no.19, 191102 (2017)
doi:10.1103/PhysRevLett.118.191102
[arXiv:1610.03071 [astro-ph.HE]].
%265 citations counted in INSPIRE as of 08 Oct 2024

%\cite{Cui:2018klo}
\bibitem{Cui:2018klo}
M.~Y.~Cui, X.~Pan, Q.~Yuan, Y.~Z.~Fan and H.~S.~Zong,
%``Revisit of cosmic ray antiprotons from dark matter annihilation with updated constraints on the background model from AMS-02 and collider data,''
JCAP \textbf{06}, 024 (2018)
doi:10.1088/1475-7516/2018/06/024
[arXiv:1803.02163 [astro-ph.HE]].
%42 citations counted in INSPIRE as of 08 Oct 2024

%\cite{Cholis:2019ejx}
\bibitem{Cholis:2019ejx}
I.~Cholis, T.~Linden and D.~Hooper,
%``A Robust Excess in the Cosmic-Ray Antiproton Spectrum: Implications for Annihilating Dark Matter,''
Phys. Rev. D \textbf{99}, no.10, 103026 (2019)
doi:10.1103/PhysRevD.99.103026
[arXiv:1903.02549 [astro-ph.HE]].
%164 citations counted in INSPIRE as of 08 Oct 2024

%\cite{ForemanMackey:2012ig}
\bibitem{ForemanMackey:2012ig}
D.~Foreman-Mackey, D.~W.~Hogg, D.~Lang and J.~Goodman,
%``emcee: The MCMC Hammer,''
Publ. Astron. Soc. Pac. \textbf{125}, 306-312 (2013)
doi:10.1086/670067
[arXiv:1202.3665 [astro-ph.IM]].
%4425 citations counted in INSPIRE as of 27 Sep 2024

%\cite{PandaX-4T:2021bab}
\bibitem{PandaX-4T:2021bab}
Y.~Meng \textit{et al.} [PandaX-4T],
%``Dark Matter Search Results from the PandaX-4T Commissioning Run,''
Phys. Rev. Lett. \textbf{127}, no.26, 261802 (2021)
doi:10.1103/PhysRevLett.127.261802
[arXiv:2107.13438 [hep-ex]].
%472 citations counted in INSPIRE as of 27 Sep 2024

%\cite{LZ:2022ufs}
\bibitem{LZ:2022ufs}
J.~Aalbers \textit{et al.} [LZ],
%``First Dark Matter Search Results from the LUX-ZEPLIN (LZ) Experiment,''
Phys. Rev. Lett. \textbf{131}, no.4, 041002 (2023)
doi:10.1103/PhysRevLett.131.041002
[arXiv:2207.03764 [hep-ex]].
%694 citations counted in INSPIRE as of 27 Sep 2024

%\cite{XENON:2020kmp}
\bibitem{XENON:2020kmp}
E.~Aprile \textit{et al.} [XENON],
%``Projected WIMP sensitivity of the XENONnT dark matter experiment,''
JCAP \textbf{11}, 031 (2020)
doi:10.1088/1475-7516/2020/11/031
[arXiv:2007.08796 [physics.ins-det]].
%380 citations counted in INSPIRE as of 25 Sep 2024

%\cite{LZ:2015kxe}
\bibitem{LZ:2015kxe}
D.~S.~Akerib \textit{et al.} [LZ],
%``LUX-ZEPLIN (LZ) Conceptual Design Report,''
[arXiv:1509.02910 [physics.ins-det]].
%377 citations counted in INSPIRE as of 08 Oct 2024

%\cite{Schumann:2015cpa}
\bibitem{Schumann:2015cpa}
M.~Schumann, L.~Baudis, L.~B\"utikofer, A.~Kish and M.~Selvi,
%``Dark matter sensitivity of multi-ton liquid xenon detectors,''
JCAP \textbf{10}, 016 (2015)
doi:10.1088/1475-7516/2015/10/016
[arXiv:1506.08309 [physics.ins-det]].
%125 citations counted in INSPIRE as of 25 Sep 2024

%\cite{Zhu:2022tpr}
\bibitem{Zhu:2022tpr}
C.~R.~Zhu, M.~Y.~Cui, Z.~Q.~Xia, Z.~H.~Yu, X.~Huang, Q.~Yuan and Y.~Z.~Fan,
%``Explaining the GeV Antiproton Excess, GeV \ensuremath{\gamma}-Ray Excess, and W-Boson Mass Anomaly in an Inert Two Higgs Doublet Model,''
Phys. Rev. Lett. \textbf{129}, no.23, 23 (2022)
doi:10.1103/PhysRevLett.129.231101
[arXiv:2204.03767 [astro-ph.HE]].
%74 citations counted in INSPIRE as of 25 Sep 2024

%\cite{ATLAS:2019wgx}
\bibitem{ATLAS:2019wgx}
G.~Aad \textit{et al.} [ATLAS],
%``Search for chargino-neutralino production with mass splittings near the electroweak scale in three-lepton final states in $\sqrt {s}$=13  TeV $pp$ collisions with the ATLAS detector,''
Phys. Rev. D \textbf{101}, no.7, 072001 (2020)
doi:10.1103/PhysRevD.101.072001
[arXiv:1912.08479 [hep-ex]].
%104 citations counted in INSPIRE as of 25 Sep 2024

%\cite{MuonCollider:2022xlm}
\bibitem{MuonCollider:2022xlm}
J.~de Blas \textit{et al.} [Muon Collider],
%``The physics case of a 3 TeV muon collider stage,''
[arXiv:2203.07261 [hep-ph]].
%103 citations counted in INSPIRE as of 26 Sep 2024

%\cite{Black:2022cth}
\bibitem{Black:2022cth}
K.~M.~Black, S.~Jindariani, D.~Li, F.~Maltoni, P.~Meade, D.~Stratakis, D.~Acosta, R.~Agarwal, K.~Agashe and C.~Aim\`e, \textit{et al.}
%``Muon Collider Forum report,''
JINST \textbf{19}, no.02, T02015 (2024)
doi:10.1088/1748-0221/19/02/T02015
[arXiv:2209.01318 [hep-ex]].
%76 citations counted in INSPIRE as of 27 Sep 2024

%\cite{Bissmann:2020lge}
\bibitem{Bissmann:2020lge}
S.~Bi\ss{}mann, G.~Hiller, C.~Hormigos-Feliu and D.~F.~Litim,
%``Multi-lepton signatures of vector-like leptons with flavor,''
Eur. Phys. J. C \textbf{81}, no.2, 101 (2021)
doi:10.1140/epjc/s10052-021-08886-3
[arXiv:2011.12964 [hep-ph]].
%43 citations counted in INSPIRE as of 25 Sep 2024

%\cite{ATLAS:2021yyr}
\bibitem{ATLAS:2021yyr}
G.~Aad \textit{et al.} [ATLAS],
%``Search for supersymmetry in events with four or more charged leptons in 139 fb$^{−1}$ of $ \sqrt{s} $ = 13 TeV pp collisions with the ATLAS detector,''
JHEP \textbf{07}, 167 (2021)
doi:10.1007/JHEP07(2021)167
[arXiv:2103.11684 [hep-ex]].
%68 citations counted in INSPIRE as of 25 Sep 2024

%\cite{Lu:2021vcp}
\bibitem{Lu:2021vcp}
C.~T.~Lu, R.~Ramos and Y.~L.~S.~Tsai,
%``Shedding light on dark matter with recent muon (g \ensuremath{-} 2) and Higgs exotic decay measurements,''
JHEP \textbf{08}, 073 (2021)
doi:10.1007/JHEP08(2021)073
[arXiv:2104.04503 [hep-ph]].
%13 citations counted in INSPIRE as of 25 Sep 2024

%\cite{Blinov:2015vma}
\bibitem{Blinov:2015vma}
N.~Blinov, S.~Profumo and T.~Stefaniak,
%``The Electroweak Phase Transition in the Inert Doublet Model,''
JCAP \textbf{07}, 028 (2015)
doi:10.1088/1475-7516/2015/07/028
[arXiv:1504.05949 [hep-ph]].
%77 citations counted in INSPIRE as of 25 Sep 2024

%\cite{Bell:2019mbn}
\bibitem{Bell:2019mbn}
N.~F.~Bell, M.~J.~Dolan, L.~S.~Friedrich, M.~J.~Ramsey-Musolf and R.~R.~Volkas,
%``Electroweak Baryogenesis with Vector-like Leptons and Scalar Singlets,''
JHEP \textbf{09}, 012 (2019)
doi:10.1007/JHEP09(2019)012
[arXiv:1903.11255 [hep-ph]].
%28 citations counted in INSPIRE as of 25 Sep 2024
\end{thebibliography}
\end{document}